\documentclass[aps,prd,twocolumn,nofootinbib,groupedaddress]{revtex4-2}

\usepackage{xcolor}
\usepackage{graphicx}
\usepackage[colorlinks=true,allcolors=blue]{hyperref}
\usepackage{aas_macros}
\usepackage{amsmath}
\usepackage{enumitem}
\usepackage[flushleft]{threeparttable}     
\usepackage[normalem]{ulem}

\def\spose#1{\hbox to 0pt{#1\hss}}
\def\lta{\mathrel{\spose{\lower 3pt\hbox{$\mathchar"218$}}
     \raise 2.0pt\hbox{$\mathchar"13C$}}}
\def\gta{\mathrel{\spose{\lower 3pt\hbox{$\mathchar"218$}}
     \raise 2.0pt\hbox{$\mathchar"13E$}}}
\def\HI{\hbox{\rm H~$\scriptstyle\rm I$}}

\def\HeI{\hbox{He~$\scriptstyle\rm I$}}
\def\HeII{\hbox{He~$\scriptstyle\rm II$}}

\def\Lya{\hbox{\rm Lyman-$\alpha$}}

\def\skm{\hbox{\rm $\mathrm {s \, km^{-1}}$}} 
\def\mwdm{\hbox{$m_{\rm{WDM}}$}}
\def\pk{\hbox{$P(k)$}}
\def\taueff{\hbox{$\tau_{\mathrm{eff}}$~}}
\def\alphae{\hbox{$\alpha_\mathrm{E}$}}
\def\Fmean{\hbox{$\langle F \rangle$}}

\newcommand{\orcidicon}[1]{\href{https://orcid.org/#1}{\includegraphics[height=\fontcharht\font`\B]{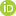}}}

\begin{document}

\title{New Constraints on Warm Dark Matter from the Lyman-$\alpha$ Forest Power Spectrum}

\author{Bruno Villasenor\,\orcidicon{0000-0002-7460-8129}}
\email{brvillas@ucsc.edu}
\affiliation{Department of Astronomy and Astrophysics, University of California, Santa Cruz, 1156 High Street, Santa Cruz, CA 95064 USA}
\author{Brant Robertson\,\orcidicon{0000-0002-4271-0364}}
\affiliation{Department of Astronomy and Astrophysics, University of California, Santa Cruz, 1156 High Street, Santa Cruz, CA 95064 USA}
\author{Piero Madau\,\orcidicon{0000-0002-6336-3293}}
\affiliation{Department of Astronomy and Astrophysics, University of California, Santa Cruz, 1156 High Street, Santa Cruz, CA 95064 USA}
\author{Evan Schneider\,\orcidicon{0000-0001-9735-7484}}
\affiliation{Department of Physics and Astronomy \& Pittsburgh Particle Physics, Astrophysics, and Cosmology Center (PITT PACC), University of
Pittsburgh, Pittsburgh, PA 15260, USA}

\date{\today}

\begin{abstract}
The forest of \Lya\ absorption lines detected in the spectra of distant quasars encodes information on the nature and properties of 
dark matter and the thermodynamics of diffuse baryonic material. Its main observable -- the 1D flux power spectrum (FPS) -- should exhibit a 
suppression on small scales and an enhancement on large scales in warm dark matter (WDM) cosmologies compared to standard $\Lambda$CDM. Here, 
we present an unprecedented suite of 1080 high-resolution cosmological hydrodynamical simulations run with the Graphics Processing Unit-accelerated code {\sc Cholla} to study the evolution of the \Lya\ forest under a wide range of physically-motivated gas thermal histories along 
with different free-streaming lengths of WDM thermal relics in the early Universe. A statistical comparison 
of synthetic data with the forest FPS measured down to the smallest velocity scales ever probed at redshifts $4.0\lta z\lta 5.2$ \cite[][]{boera2019a} yields a lower limit $m_{\rm WDM}>3.1$ keV (95 percent CL) for the WDM particle mass and constrains the amplitude and spectrum of the 
photoheating and photoionizing background produced by star-forming galaxies and active galactic nuclei at these redshifts. Interestingly, our Bayesian inference analysis appears to weakly favor WDM models
with a peak likelihood value at the thermal relic mass of $m_{\rm WDM}=4.5$ keV.
We find that the suppression of the FPS from free-streaming saturates at $k\gta 0.1\,\skm$ because of peculiar velocity smearing, and this saturated suppression combined with a slightly lower gas  temperature provides a moderately better fit to the observed small-scale FPS for WDM cosmologies.
\end{abstract}


\maketitle

%
%
\section{Introduction}

The $\Lambda$-cold dark matter ($\Lambda$CDM) cosmological paradigm has been immensely successful at matching across cosmic time 
observations spanning physical scales from the horizon length \cite[][]{planck2020} all the way down to galaxy scales \cite[e.g.,][]{chabanier2019, Colin+2000, Avila+2001}, 
and a vast menagerie of hypothetical non-baryonic elementary particles has been proposed to explain 
the astrophysical data \cite[][]{feng2010}.  Cold dark matter particles have negligible thermal velocities in the matter-dominated era and 
therefore clump gravitationally even on the smallest sub-galactic scales, a property that has caused persistent challenges with observations 
of the abundances and density profiles of dwarf galaxies in the local Universe \cite[e.g.,][]{bullock2017}. Warm dark matter (WDM) is a 
simple modification of CDM that has been proposed to suppress small-scale power and alleviate some of these 
problems \cite[][]{bode2001}. WDM particles of a few keV have significant intrinsic velocities from having decoupled as thermal relics or 
been produced by non-equilibrium processes, and one of the effects of their Mpc-range {\it free-streaming length} is to limit the gravitational collapse of structures and produce a cut-off in the matter power spectrum. 

Intergalactic hydrogen at redshift $2\lta z \lta 5$ scatters \Lya\ radiation and produces absorption features in the spectra of 
distant quasars. This ``\Lya\ forest" is a powerful cosmological probe as it traces density fluctuations, the underlying dark matter 
web-like distribution (the ``cosmic web"), and the ionization state and temperature of the intergalactic medium (IGM) at scales and 
redshifts that cannot be probed by any other observable \cite[][]{hernquist1996a,croft1998,meiksin2009,mcquinn2016a}. The primary statistic 
derived from spectroscopic data is the 1D power spectrum  (FPS) of the flux distribution in the forest -- the 
Fourier transform of the fractional flux autocorrelation function in velocity space --  which arguably provides the best tool for 
distinguishing between CDM and WDM models \cite[][]{seljak2006,viel2013a,baur2016,garzilli2019,garzilli2021}. 
The FPS is observationally-accessible over 
a wide range of redshifts, involves the fundamental, well-known physics of the hydrogen atom, and is largely free from the uncertain baryonic physics (star formation, feedback, and metal cooling) that affects, e.g.,  the abundance of Milky Way satellites and the central density cores of dwarf galaxies \citep{Bullock+2017}.
Interpreting such observations requires 
expensive hydrodynamical simulations of the IGM that cover an extensive range of uncertain IGM photoionization and photoheating histories and 
consistently maintain high resolution throughout a statistically representative sub-volume of the Universe, a traditional limiting factor 
in previous analyses \cite[e.g.,][]{viel2013a,lukic2015a,Irsic+2017b,bolton2017,walther2021}. At present, the tightest lower limit 
to the mass of a thermal WDM relic, $m_{\rm WDM}>1.9$ keV (95 percent CL), is obtained for highly conservative thermal histories of 
intergalactic gas and without a marginalization over the properties of the cosmic ionizing background
\cite[][]{garzilli2021}.

In this Paper, we revisit these constraints in light of recent measurements with the {\it Keck} and {\it VLT} telescopes of the \Lya\ FPS down 
to the smallest scales ever probed at redshifts $4\lta z\lta 5$ \cite[][]{boera2019a}, using a massive suite of 1080 high-resolution cosmological 
hydrodynamics simulations that are part of the {\sc Cholla} IGM Photoheating Simulations (CHIPS) suite \cite[][]{villasenor2021a, villasenor2022}. High-redshift observations provide better limits on the mass of a WDM particle, as free-streaming becomes more prominent in velocity space and the effect of the non-linear evolution of the matter density field -- resulting in a 
transfer of power from large scales to small scales -- is weaker at these epochs. There is also observational
evidence for a local minimum of the temperature of the IGM -- and therefore a corresponding minimum in the thermal cut-off 
length scale -- in the redshift range $4 < z < 5$ \cite[][]{villasenor2022,gaikwad2020a}, making this era an optimal epoch 
for deriving bounds on dark matter properties.

\begin{figure*}[htb!]
    \centering
    \includegraphics[width=\linewidth]{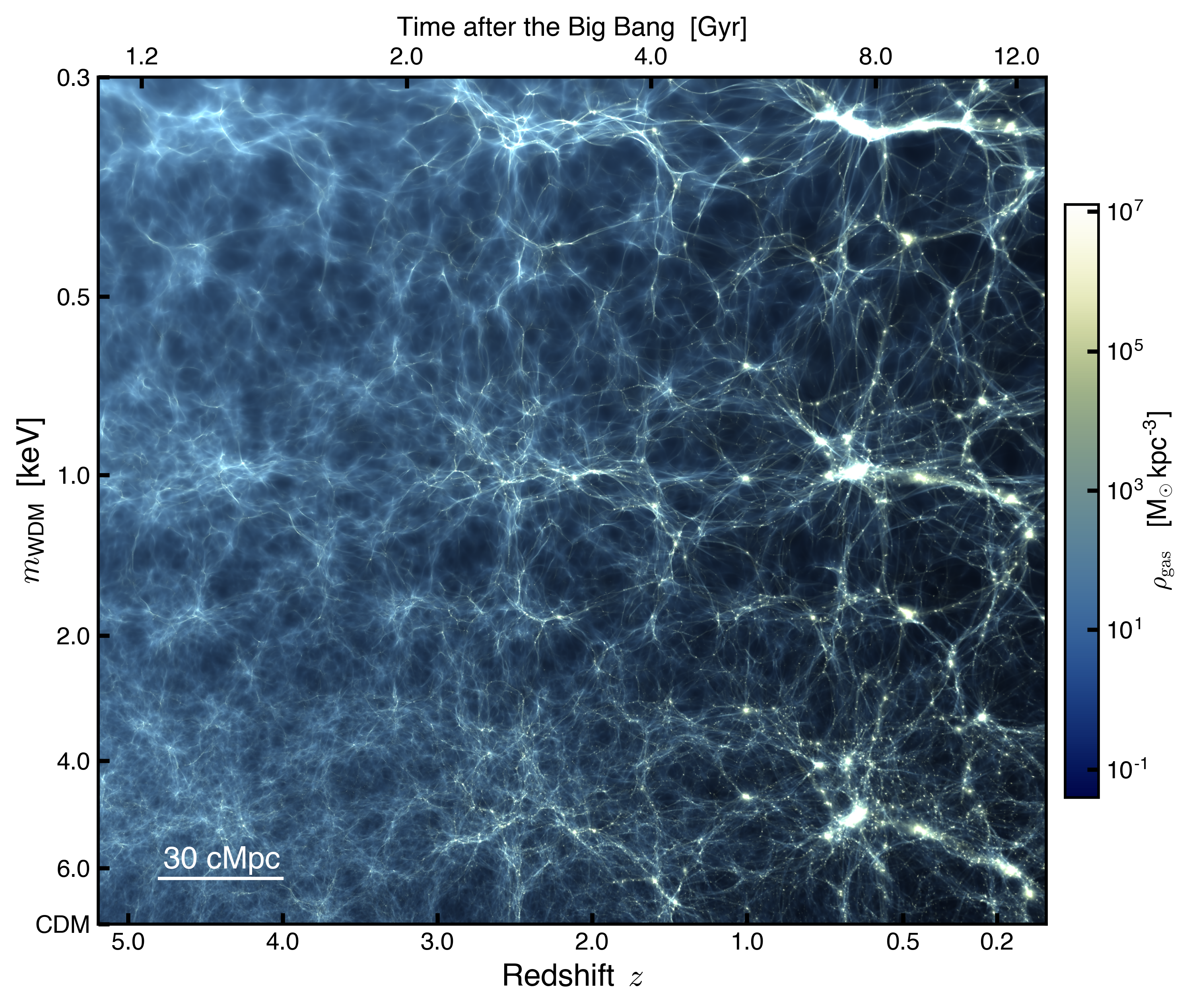}
    \caption{Impact of particle free-streaming on baryonic structures at $0<z<5.2$. The evolution of the gas density from a slice through the IGM was obtained from a subset of 8 CHIPS simulations where the mass of the warm dark matter particle \mwdm\ was increased from 0.3 keV to $\infty$. All simulations assume the same gas thermal history from the best-fit model presented in \cite{villasenor2022}. Due to thermal pressure, the gas distribution is smoothed relative to the dark matter.
    }
    \label{fig:wdm_dens_slice}
\end{figure*}

\section{Basic equations and characteristic scales}

The hydrogen and helium photoheating ${\cal H}_i$ and photoionization $\Gamma_i$ rates per atom $i$ depend on the intensity of a uniform UV background radiation field $J(\nu,z)$ as

\begin{eqnarray}
\Gamma_i(z) &  =  & 4\pi \int_{\nu_i}^{\infty} \frac{J(\nu,z)}{h\nu} \sigma_i(\nu) d\nu, \\
{\cal H}_i(z) & = & 4\pi \int_{\nu_i}^{\infty} \frac{J(\nu,z)}{h\nu}\left(h\nu-h\nu_i\right) \sigma_i(\nu) d \nu,
\label{eq:photoionization_photoheating}
\end{eqnarray}
where $\nu_i$ and $\sigma_i(\nu)$ are the threshold  frequency and  photoionization cross-section, respectively. The total photoheating rate ${\cal H}$ is summed over the species $i=\,$\HI, \HeI, and \HeII\ each of proper number density $n_i$, ${\cal H}=\sum_i n_i {\cal H}_i$. These rates, together with radiative recombinations, adiabatic compression, Compton and expansion cooling, and the gas density and peculiar velocity fields shaped by gravity and by pressure forces, fully determine the hydrogen \Lya\ absorption optical depth in velocity space $v$ \cite{hui97},

\begin{equation}
\tau_{v}={\sigma_\alpha c}\,\int \frac{n_{\rm HI}}{\sqrt{\pi} b H(z)}\, e^{-(v-u-v_p)^2/b^2} du. 
\label{eq:tau}
\end{equation}
Here, $\sigma_\alpha=\pi e^{2} f_{12}/(m_{e} c 
\,\nu_\alpha)$, $\nu_\alpha$ and $f_{12}$ are the frequency and upward oscillator strength of the \Lya\ transition, $b=(2k_BT/m_{\rm H})^{1/2}$ is the 
Doppler parameter, $T$ the gas temperature, 
$H(z)$ the Hubble parameter, $u$ and $v_p$ 
are, respectively, the real-space coordinate (in km s$^{-1}$) and the peculiar velocity along the line of sight, and we have assumed a Gaussian profile.
Denoting with $\bar z$ some mean redshift of interest, e.g. the redshift of a simulation output or the average redshift of any given data subset, one can write

\begin{equation}
u(x)= {H(\bar z)\over (1+\bar z)}(x-\bar x),
\label{eq:uvel}
\end{equation}
where $\bar x$ is the comoving position at which the redshift from Hubble expansion is exactly $\bar z$, and $H(\bar z)$ is the Hubble expansion rate. Line center in the rest frame of intervening hydrogen occurs at velocity $v=u+v_p$, which is related to the observed frequency $\nu_0$ by

\begin{equation}
\nu_\alpha=\nu_0(1+\bar z)\left(1+{v\over c}\right). 
\end{equation}
Because of peculiar velocities, a photon observed at $\nu_0$ can have the same rest-frame frequency $\nu_\alpha$  at more than one place in its trajectory from the quasar to the observer. 

In practice, only a limited range of $u$ values contributes to $\tau_v$, and one can replace in Equation (\ref{eq:tau}) $H(z)$ with $H(\bar z)$. 
We can then define the flux contrast 
$\delta_F(v)=F(v)/ \Fmean -1$, where $F(v)=\exp(-\tau_v)$ is the flux at velocity $v$ and \Fmean\ is the mean transmitted flux at a given redshift, and decompose each absorption spectrum into Fourier modes $\tilde \delta_F(k)$. Their variance as a function of the Fourier wavenumber $k=2\pi/v$ is the FPS over some velocity interval $\Delta v$, 
\begin{equation}
P_F(k)=\Delta v\langle \tilde \delta_F(k)^2\rangle,
\label{eq:Pk}
\end{equation}
which is commonly expressed in terms of the dimensionless quantity $\Delta_F^2(k)=kP_F(k)/\pi$.

Four physical effects (cf. \citep{garzilli2019})
act to erase small-scale power in the FPS:

\begin{enumerate}[leftmargin=*]

\item {\bf Doppler broadening caused by gas thermal velocities along the line of sight.} Assuming a Gaussian smoothing kernel in the $\tilde 
\delta_F(k)$ field of the form $\exp(-k^2\sigma_{\rm th}^2/2)$, where $\sigma_{\rm th}=(k_BT/m_{\rm H})^{1/2}$ is the broadening velocity scale, a cut-off in the FPS arises  then at the proper wavenumber 
\begin{equation}
k_{\rm th}={1\over \sigma_{\rm th}}=0.11\,T_4^{-1/2}\,{\rm km^{-1}\,s},
\label{eq:kth}
\end{equation}
where $T_4\equiv T/10^4\,{\rm K}$. 

\item {\bf Pressure support.} Pressure smooths
gas density fluctuations in the {\it spatial} Fourier domain with a Gaussian kernel \citep{gnedin1998,gnedin2003} 
\begin{equation}
W(\tilde k, \tilde k_F)=\exp(-\tilde k^2/\tilde k_F^2),
\end{equation}
where $\tilde k$ is a comoving wavenumber,  
\begin{equation}
\tilde k_F=g\tilde k_J=g\,{a\over c_s} 
\sqrt{4\pi G\bar \rho}
\end{equation}
is the `filtering scale', and 
$\tilde k_J$ is the wavenumber corresponding to the Jeans scale. Here, 
$c_s=\left[5k_BT/(3\mu m_{\rm H})\right]^{1/2}$ is the sound speed, $\mu$ is the mean molecular weight ($\mu=0.61$ for an admixture of ionized hydrogen and singly ionized helium), 
$\bar \rho$ is the mean total (dark matter plus baryons) mass density of the Universe, and $a$ is the scale factor. The $g$ factor accounts for the fact that the filtering scale depends on the prior thermal history of the IGM, and not just the instantaneous Jeans scale. Typically, $g>1$ after reionization \citep{gnedin2003}.

Mapping the 3D filtering scale to a broadening velocity 
scale $\sigma_F=\sqrt{2}\,Ha/\tilde k_F$ along the line of sight results in a cut-off of the FPS at proper wavenumber
\begin{equation}
k_F={1\over \sigma_F}\simeq {\sqrt{3}g\over 
2 c_s}={0.06\,g}\,T_4^{-1/2}\,{\rm km^{-1}\,s},
\label{eq:kF}
\end{equation}
where the second equality is valid at high enough redshifts. Under the physically-motivated assumptions that the gas temperature decays as $1/a$ after reionization completes around redshift $z\sim 6$, the factor $g\sim4-10$
at the redshifts $4\lta z\lta 5.2$ of interest here \cite{gnedin1998}.
Pressure filtering, always lagging behind the growth of the Jeans length, is therefore subdominant compared to the thermal broadening of the absorption spectrum \cite{peeples10,kulkarni15,nasir16,boera2019a}.

\item {\bf Non-linear peculiar velocities in the gas.} Peculiar motions smear
power along line-of-sight (the ``fingers-of-God" effect")
over a range of scales comparable with the one-dimensional velocity dispersion. A fit to the results of numerical simulations with a Gaussian 
in Fourier space yields \citep{gnedin2002} 
\begin{equation}
\sigma_k=0.25~{\rm km^{-1}\,s} \left({k\over {\rm km^{-1}\,s}}
\right)^{1/2}. 
\end{equation}
Setting $k=\sqrt{2}\sigma_k$, a cut-off in the FPS arises at proper wavenumber
\begin{equation}
k_v= 0.13\,{\rm km^{-1}\,s}. 
\end{equation}

\item {\bf Dark matter free-streaming.} This effect can be described by the wavenumber, $\tilde k_{1/2}$, at which the WDM matter power spectrum is suppressed relative to 
the CDM case by a factor of 2 \cite{viel2005},
\begin{equation}
\tilde k_{1/2}=4.25\,\left({\mwdm\over 1\,{\rm keV}}\right)^{1.11}\,{\rm Mpc^{-1}}
\end{equation}
for a standard thermal relic. The conversion to a proper wavenumber in inverse km s$^{-1}$, $k_{\rm FS}=\tilde k_{1/2}/(aH)$, yields a cut-off in the FPS at
\begin{equation}
k_{\rm FS}=0.05\left({\mwdm\over 1\,{\rm keV}}\right)^{1.11}\left({6\over 1+z}\right)^{1/2}\,{\rm km^{-1}\,s}.
\label{eq:kFS}
\end{equation}
Note how the free-streaming cut-off moves toward larger scales (smaller $k$) at higher redshift.
\end{enumerate}

The numerical values of $k_{\rm th}, k_F,$ $k_v$, and 
$k_{\rm FS}$ above suggest that free-streaming, thermal and velocity broadening effects all set in at similar scales. Thus, comparing the observations of the FPS to a large set of models that accurately depict the suppression on the FPS in the non-linear regime associated to each of these mechanisms is critical to constrain the WDM particle mass. 
In the following section, we present our grid of simulations that evolve a total of 1080 models, which simultaneously vary the impact of free-streaming, Doppler broadening, and pressure smoothing. A comparison between  our grid of models and the observed FPS will provide new constraints on the mass of the dark matter particle that, for the first time, marginalize over all the important confounding physical effects.

\begin{figure*}[htb!]
    \centering
    \includegraphics[width=\linewidth]{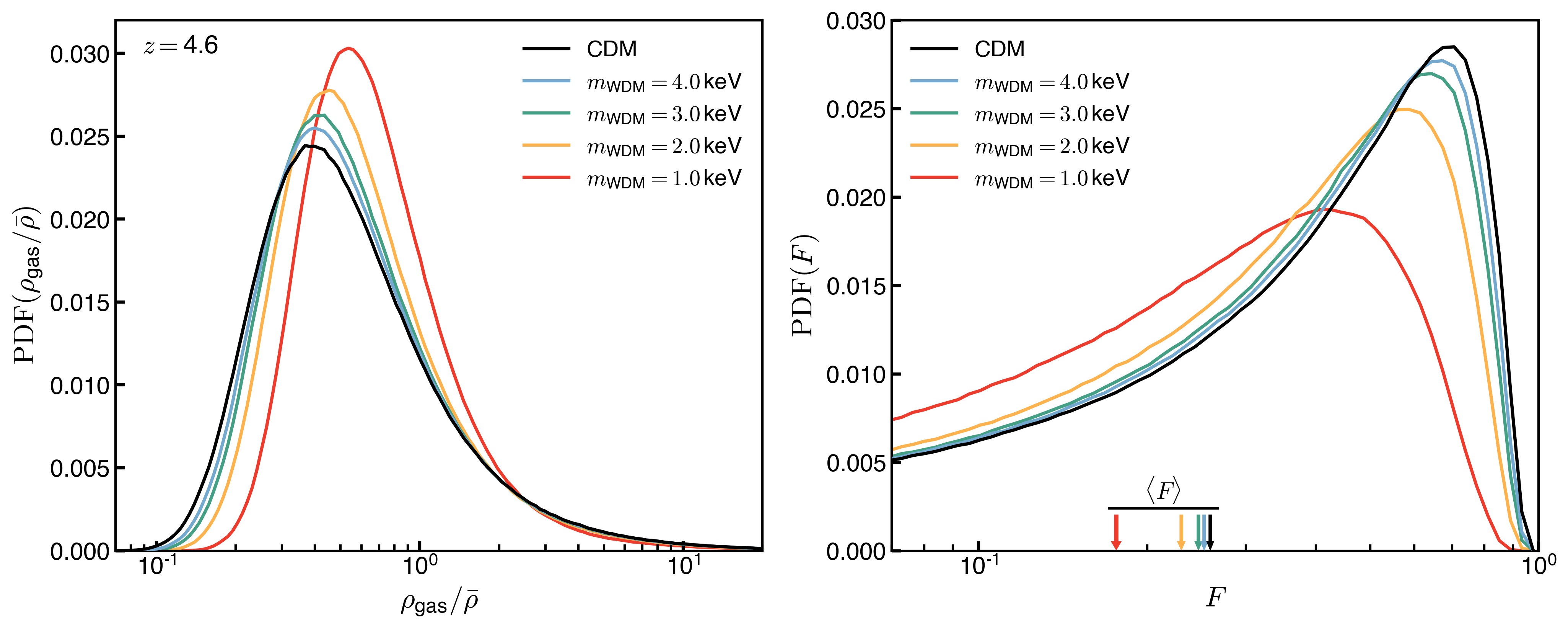}
    \caption{ {\it Left:} Volume-weighted probability distribution function of the gas density at $z=4.6$ in WDM cosmologies compared to $\Lambda$CDM.
    Due to free-streaming, gas that otherwise would collapse into small-scale structures redistributes on larger scales. This suppressed collapse shifts the 
    peak of the distribution closer to the mean density  $\bar{\rho}$.   
    {\it Right:} Gas redistribution in WDM cosmologies results in an increase of the \Lya\ scattering optical depth of the IGM and a decrease of the transmitted flux $F$. Arrows display the decrease of the mean transmission \Fmean\ as \mwdm\ decreases. 
    }
    \label{fig:dens_flux_distributions}
\end{figure*}

%
%
\section{Simulations}

The simulations used in this work were performed using the GPU-native MPI-parallelized code {\sc Cholla} \cite[][]{schneider2015a}, 
and evolve the equations of hydrodynamics on a uniform Cartesian grid while simultaneously tracking the non-equilibrium ionization 
states of hydrogen and helium using the GRACKLE library \cite[][]{smith2017a}.  A spatially uniform, time-dependent UV radiation 
background was assumed in the form of redshift-dependent photoionization 
and photoheating rates per ion, $\Gamma_i$ and ${H}_i$, for the species \HI, \HeI, and \HeII. The initial conditions at $z=100$ were generated using the 
MUSIC code \cite[][]{Hahn+2011_Music} for a flat cosmology with parameters 
$h = 0.6766$,
$\Omega_M = 0.3111$, 
$\Omega_{\Lambda} = 0.6889$, $\Omega_b = 0.0497$, $\sigma_8 = 0.8102$, and $n_s = 0.9665$, consistent with constraints from Planck data \cite{planck2020}. 
The initial conditions for all runs were generated from identical random number seeds to preserve the same amplitude and phase for all initial Fourier 
modes across the simulation suite. The volume and numerical size of the simulations correspond to $L$ = 25 $h^{-1}$Mpc and $N$ = 2$\times$1024$^3$ 
cells and particles. All simulations
were run on the {\it Summit} computing system at the Oak Ridge National Laboratory. 
The effect of a thermally produced WDM particle was introduced by modifying the input transfer function following the fitting formula of \cite{viel2005}, and no attempt was made to explicitly incorporate particle thermal velocities in the initial conditions \cite[][]{maccio2013}. WDM thermal velocities will be only a few percent of the velocities due to gravitational acceleration at the cosmic time of our initial conditions, and should not be important on the scales resolved by our simulations. 

To compare the properties of the IGM in our simulations to observations, we extracted synthetic \Lya\ forest spectra measuring \HI\ densities, temperatures, and peculiar velocities of the gas using 4096 skewers randomly distributed for each simulation along each axis of the box. The optical depth $\tau_v$ along each discretized line of sight was estimated as described in \cite{villasenor2021a}. From the set of lines of sight, we computed the transmitted flux and the mean one-dimensional power spectrum of the fractional transmission (see \S5.4 in \cite{villasenor2021a} for a detailed description).

\subsection{WDM Effects on the IGM}

The impact of particle free-streaming on the gas density structure of the IGM at fixed thermal history is shown in Figure \ref{fig:wdm_dens_slice}. The image displays a slice through the gaseous cosmic web at $0\le z \le 5.2$ generated from a set of 8 simulations that vary the mass of the dark matter particle \mwdm\ increasing from 0.3 keV to $\infty$. The simulations follow the thermal history produced by the baseline photoheating and photoionization history presented in \cite{villasenor2022}. 

The main effect of the free-streaming of keV particles on the matter density distribution is to suppress the formation of small-scale structure, reducing the clumpiness of the cosmic web. As free-streaming washes out small-scale inhomogeneities and decreases the concentration of overdense regions, baryon redistribution leaves an excess (relative to CDM) of close-to-mean density gas to permeate the IGM.
Figure \ref{fig:dens_flux_distributions} (left)
shows the change in the volume-weighted probability distribution function of the gas density at $z = 4.6$ for different \mwdm\ cosmologies compared to a CDM simulation with the same thermal history.
The enhanced fraction of close-to-mean density as a 
consequence of free-streaming increases the opacity of the IGM to \Lya\ scattering. Figure \ref{fig:dens_flux_distributions} (right) shows the probability distribution function of the \Lya\ transmitted flux $F$ along the skewer set for the WDM and CDM runs
at $z = 4.6$. The distribution of transmitted flux shifts to lower values with increasing free-streaming lengths.
The arrows at the bottom of the figure display the decrease of the mean transmission \Fmean\ of the forest as \mwdm\ decreases.  

While free-streaming (in the regime $\mwdm \geq 1\,$ keV) changes the IGM opacity it has a negligible effect on its density-temperature relation, often approximated with a
power-law model $T=T_0 (\rho / \overline{\rho})^{\gamma -1}$,
and the parameters $T_0$ and $\gamma$ 
are virtually unchanged -- for fixed photoionization and photoheating histories -- from the CDM case.

\subsection{WDM Effects on the Flux Power Spectrum}

\begin{figure}[htb!]
    \centering
    \includegraphics[width=\linewidth]{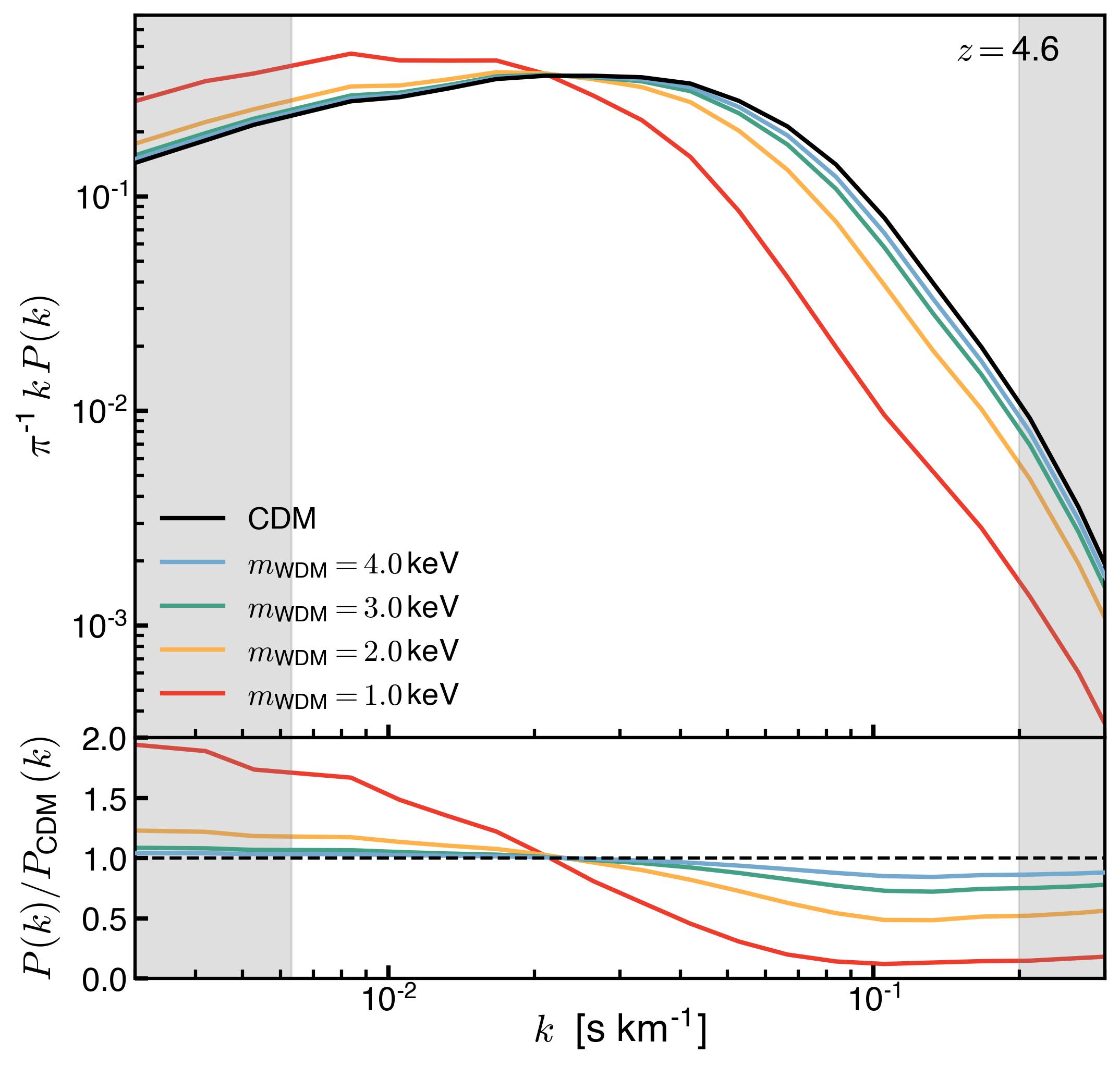}
    \caption{Transmitted FPS at $z=4.6$ from simulations that vary the WDM particle mass at fixed thermal history. The bottom panel displays the fractional difference relative to the CDM case. The suppression of small-scale structure due to free-streaming results in reduced power compared to CDM for $ k \gtrsim 0.02 \,\, \mathrm{ s \, km^{-1}}$. On large scales, the increase of the overall normalization compared to CDM is associated with the excess of close-to-mean density  gas and resulting reduction of the mean 
    transmitted flux \Fmean.  The gray bands
    denote the regions that fall outside the observational measurements of \cite{boera2019a}.
    }
    \label{fig:ps_wdm_effects}
\end{figure}

The suppression of small-scale density fluctuations from
free-streaming translates into 
a decrease of the small-scale FPS compared to CDM. Figure \ref{fig:ps_wdm_effects} shows the dimensionless FPS of the \Lya\ transmitted flux, $\pi^{\mathrm{-1}} k P(k)$,
measured at $z=4.6$ in simulations with different WDM particle mass. The ratio of $P(k)$ with respect to the 
standard CDM case is depicted in the bottom panel. Free-streaming  affects the FPS differently on small and large scales. On small scales, $k\gtrsim 0.02~\mathrm{s\, km^{-1}}$, the suppression of density fluctuations reduces power. On large scales, 
$ k\lesssim 0.02~\mathrm{s \, km^{-1}}$, there is an increase of the overall normalization of \pk\
associated with the reduction of the mean transmitted flux \Fmean.     

\subsection{Simulation Grid}

To constrain WDM cosmologies, we have performed an unprecedented grid of 1080 high-resolution simulations 
(with $L$ = 25 $h^{-1}$Mpc and $N$ = 2$\times$1024$^3$) for a variety of particle masses \mwdm\ and  thermal histories of the IGM that alter the impact of 
free-streaming, Doppler broadening, and pressure smoothing on the forest FPS.
Here, we use the fiducial model for the hydrogen and helium photoionization and photoheating rates of \citep{villasenor2022} 
(hereafter V22) as a template to generate different gas reionization and thermal histories.
The V22 photorates are a modification of the rates by \citep{puchwein2019a} and were determined
by running hundreds of cosmological simulations to
produce a reheating history in agreement with observations of the hydrogen \Lya\ FPS as well as the effective opacity of the \HeII\ \Lya\ forest (see \citep{villasenor2022} for details).\footnote{In our modeling, 
the ionization and thermal evolution of the IGM is primarily determined by the radiation emitted by star-forming galaxies and active galactic nuclei (AGNs)  over cosmic history. The  photoionization and photoheating rates of \citep{puchwein2019a} were computed from the intensity of the UV background (UVB) radiation field, which was in turn determined by the emissivity of the radiating sources. An improved treatment of
the IGM opacity to ionizing radiation that consistently captures the transition from a neutral to an ionized IGM was also adopted in 
\citep{puchwein2019a}. This UVB model, when applied to cosmological simulations, results in a hydrogen reionization era that completes by redshift $\sim 6$ (V22).}

Our grid of models is based on three different transformations of the V22 fiducial rates: 

\begin{enumerate}[leftmargin=*]

\item We rescale the hydrogen (\HI) and helium (\HeI) photoionization rates $\Gamma_i$ by a constant factor $\beta$. This rescaling mainly impacts the neutral hydrogen (and helium) fractions, changing the mean transmitted flux \Fmean\ and therefore the overall normalization of the FPS.

\item We modify the \HI\ and \HeI\ photoheating rates ${\cal H}_i$ by rescaling the mean energy of the ejected 
photoelectrons $\mathcal{H}_i(z) / \Gamma_i(z)$ 
by a factor \alphae. This modification 
changes the temperature of the IGM and the impact of thermal Doppler broadening on 
the small-scale FPS.

\item We change the timing of hydrogen reionization by offsetting the \HI\ and \HeI\ photoionization and photoheating rates by an amount $\Delta z$ -- a value $\Delta z>0$ shifts reionization to earlier times.
This shift affects the time available for the IGM to cool by adiabatic expansion and respond to thermally induced pressure gradients, changing the impact of pressure smoothing on the small-scale FPS.

\end{enumerate}

The transformations applied to the \HI\ and \HeI\ photorates described above can be expressed as 

\begin{equation}
\begin{aligned}
\Gamma_i(z) &\rightarrow \beta \, \Gamma_i^{\mathrm{V22}}\,( z  - \Delta z), \\ 
\mathcal{H}_i(z) & \rightarrow \beta \, \alphae\, \mathcal{H}_i^{\mathrm{V22}}( z - \Delta z).
\end{aligned}
\label{eq:rates_transformation}
\end{equation}
In this work we shall compare the results of our simulations to observational determinations of the FPS in the redshift range $4.2 \leq z \leq 5.0$. In the V22 model the IGM cools roughly adiabatically after reionization 
at $z\sim6$ until AGNs begin to photoionize \HeII\ and heat intergalactic gas at $z\sim 4.5$, resulting in a local minimum of the IGM temperature at this epoch. The effect of \HeII\ reheating becomes dominant, however, only at $z<4$ and does not significantly impact the FPS at $z \geq 4.2$. Accordingly, we do not marginalize over the reionization history of helium, and fix the \HeII\ photorates to those of V22. We note that since our simulations assume a spatially-uniform UV background, there are no couplings between, e.g., the UV radiation field, the dark matter density, and peculiar motions that would affect the interpretation of our power spectrum results.

Each model in our simulation suite is therefore characterized by the four-dimensional parameter vector $\theta = \{\mwdm, \beta, \alphae, \Delta z \}$, and 
our parameter grid is shown in Table \ref{table:grid_parameters}. The range of dark matter masses were chosen to span from very light WDM ($\mwdm=1$keV) to CDM. After running some
test calculations the ranges of thermal parameters ($\alpha$, $\beta$)
were chosen to encompass the parameter space with large 
likelihood. The range of $\Delta z$ shifts span about 200 million years of cosmic time
around the end of reionization.
Full cosmological hydrodynamical simulations were performed for
all possible combinations of these four parameters, requiring
the suite of 1080 simulations used for our analysis.

\begin{table}
  \begin{threeparttable}
    \caption{WDM-CHIPS Simulation Grid\label{table:grid_parameters}}
     \begin{tabular}{cc}
        \toprule
        Parameter & Parameter Values \\ [2pt]
        \hline
        \mwdm $\,$ [keV] $\,$ & 1, 2, 3, 4, 5, 6, 8, 12, 20, 40, 80, CDM\\
        $\beta$ & 0.6, 0.8, 1.0, 1.2, 1.4, 1.8 \\
        $\alpha_E$ & 0.1, 0.5, 0.9, 1.3, 1.7 \\
        $\Delta z$ & -0.5, 0.0, 0.5 \\
        \hline
     \end{tabular}
    \begin{tablenotes}
      \small
      \item NOTE.- The grid of models consist of the 1080 possible combinations of the parameters 
           $\theta = \{ \mwdm, \beta, \alphae, \Delta z \}$. Each simulation evolves an $ L = 25 h^{-1} \mathrm{Mpc}$ and $N=2\times 1024^3$ box.
    \end{tablenotes}
  \end{threeparttable}
\end{table}

\subsection{Effects of Model Variations on the Flux Power Spectrum }
\label{sec:pk_variation}

\begin{figure*}[htb!]
    \centering
    \includegraphics[width=\linewidth]{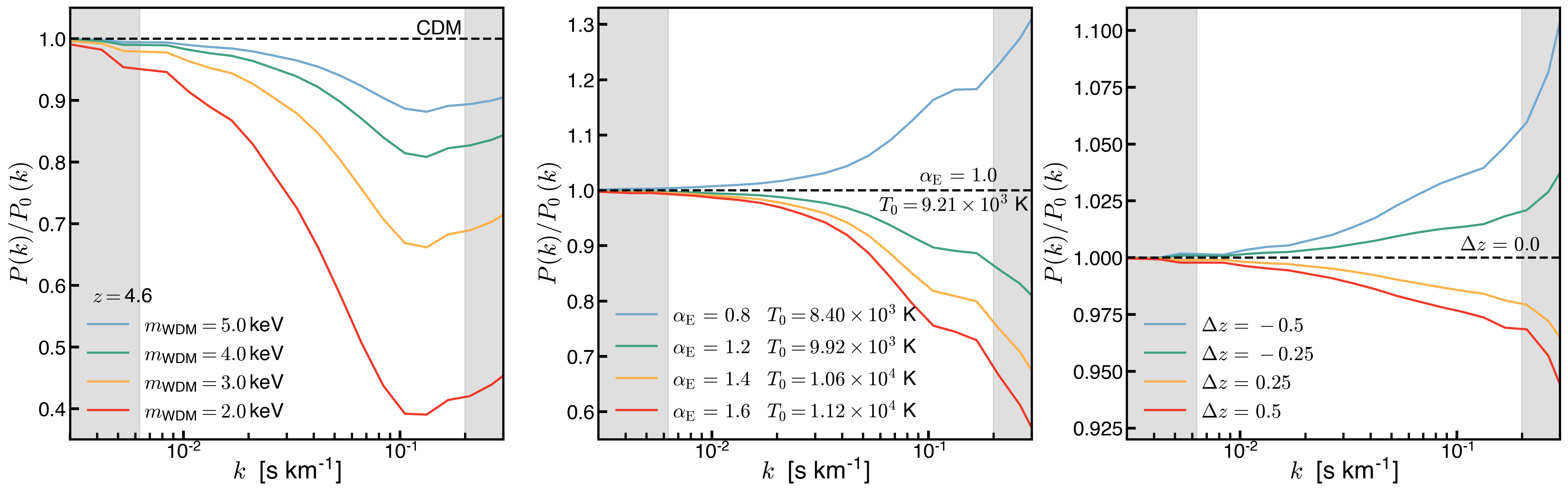}
    \caption{ Effects of free-streaming (left), thermal broadening (center) and pressure smoothing (right) on \pk\ at $z=4.6$, 
    compared with the $P_0(k)$ measured from
    a simulation that adopts the V22 fiducial model ($\Lambda$CDM, $\beta =1$, $\alphae =1$, and $\Delta z = 0$). Gray bands show the regions outside 
    the observational measurements from \cite{boera2019a}. Free-streaming has the largest effect on the shape of \pk\ relative to the fiducial CDM model, with suppression of $10-60$\% at $k\sim0.1\,\, \skm$ for $m_{\rm WDM}\sim5-2\,$keV. On smaller scales the free-streaming suppression saturates due to peculiar velocities (see text and Figure \ref{fig:ps_real_space}). Thermal broadening of lines induces a $10-20$\% effect at $k\sim 0.1\,\, \skm$ for mean IGM temperature changes of $\Delta T_0\sim1,000-2,000\,$K. Shifting the redshift of reionization by $\Delta z\sim0.5$ leads to differences in the impact of pressure smoothing of $1-2$\% at $k\sim0.1\,\,\skm$.
    }
    \label{fig:ps_effects_comparison}
\end{figure*}

Suppression of small-scale fluctuations in the \Lya\ forest can be attributed to decreased small-scale inhomogeneities in the intergalactic gas due to the 
free-streaming of  WDM particles, but also to Doppler broadening of absorption lines and pressure smoothing of gas overdensities. Providing accurate constraints on the WDM particle mass from observations of the FPS requires differentiating the impact that these processes have on \pk, and sampling over a dense set of models that simultaneously vary the effect of each mechanism.

Figure \ref{fig:ps_effects_comparison} illustrates the impact on the FPS of free-streaming (left), thermal broadening (center), and pressure smoothing (right) at $z=4.6$. Shown is the
ratio of the \pk\ measured in simulations that independently vary the model parameters \mwdm\ (left), \alphae\ (center) and $\Delta z$ (right) to the 
$P_0(k)$ 
measured from a simulation that adopts the fiducial V22 model corresponding to a $\Lambda$CDM cosmology and parameters $\beta =1$, $\alphae =1$, and $\Delta z = 0$.
To isolate small-scale effects, for this comparison the \Lya\ optical depths of the skewers from each simulation were rescaled to produce the same \Fmean\ of the fiducial model as this 
equalizes \pk\ on large scales. Additionally, for the simulations that vary $\Delta z$ (right), we separate the impact of pressure smoothing by also rescaling the instantaneous gas temperature along the skewers to have the same value at mean density $T_0$ as the fiducial model. 
Variations of the model parameter $\beta$ mainly impact the ionization fraction of intergalactic hydrogen, changing the mean transmitted flux and thereby rescaling the overall normalization of the FPS. For a detailed analysis on this effect we refer the reader to Appendix B of \cite{villasenor2022}. 

As shown in Figure \ref{fig:ps_effects_comparison} (center), the effect from increasing/decreasing the IGM temperature is to suppress/boost the FPS for 
$k\gtrsim 0.01 \, \, \skm $ relative to the fiducial model. On these scales 
\pk\ is also altered by pressure smoothing (right). An earlier (later) reionization epoch suppresses (boosts) power as overdensities have more (less) time to respond to thermally-induced pressure gradients.
Figure \ref{fig:ps_effects_comparison}
also makes clear the small-scale \pk\ behavior of
WDM models (left) relative to the fiducial CDM model.
On (quasi-)linear scales
(e.g., $k<0.1~\mathrm{s}\,\mathrm{km}^{-1}$), the effect on the power spectrum of increasing the WDM particle mass is well captured by the exponential cut-off 
of Equation (\ref{eq:kFS}).
The suppression of \pk\ due to free-streaming increases with decreasing $\mwdm$, and extends
to $k \geq 0.1 \,\, \skm$ where it saturates owing to 
peculiar motions.
We argue that this important feature could be key for
constraining the nature of dark matter.

\begin{figure}[htb!]
    \centering
    \includegraphics[width=\linewidth]{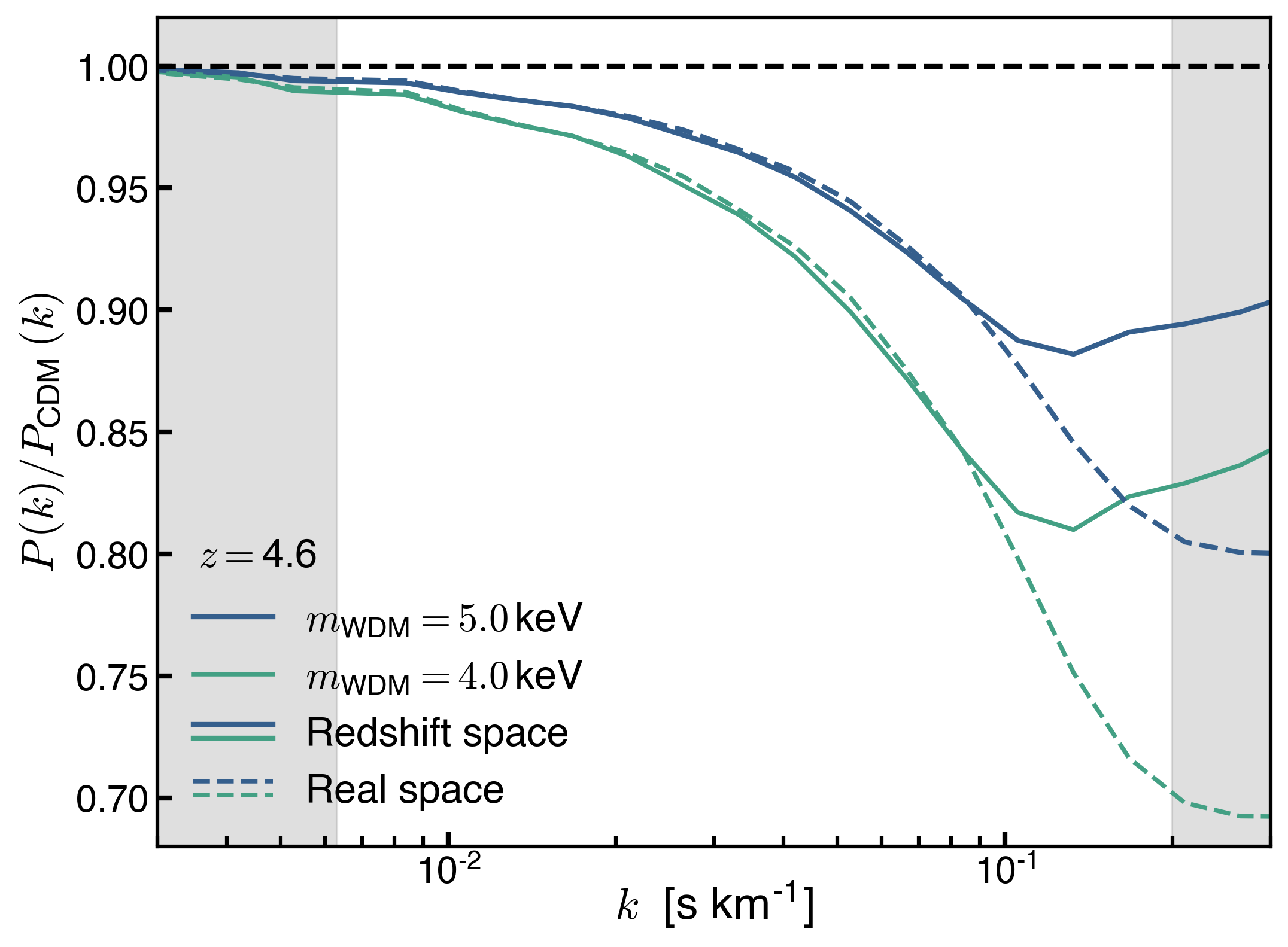}
    \caption{ Suppression of \pk\ due to free-streaming relative to CDM at $z=4.6$. The full lines show the ratio $P(k)/P_\mathrm{CDM}(k)$ where \pk\ has been measured 
    from the \Lya\ transmitted flux computed in redshift space. For the dashed lines the transmitted flux from the CDM and WDM case is computed in real space 
    (ignoring peculiar velocities). In real space the suppression of \pk\ due to free-streaming continues to $k \gtrsim 0.2 \,\, \skm$, while in redshift 
    space the suppression saturates at $k \sim 0.1 \,\, \skm$.        
    }
    \label{fig:ps_real_space}
\end{figure}

Interestingly, we find that if \pk\ is measured in configuration (`real') space instead of redshift space (ignoring gas peculiar velocities) then the suppression of \pk\ continues to $k \sim 0.2 \,\, \skm$. Figure \ref{fig:ps_real_space} displays the ratio $\pk / P_\mathrm{CDM}(k)$ for simulations with WDM particle mass $\mwdm=5 \, \mathrm{keV}$ (blue) and $\mwdm=4 \, \mathrm{keV}$ (green). Dashed lines show the same ratio $\pk / P_\mathrm{CDM}(k)$ but in this case the FPS is computed from synthetic spectra in real space. Here, the effective opacity of the skewers of the WDM simulations has been rescaled 
to match CDM on large scales.

\begin{figure*}[htb!]
    \centering
    \includegraphics[width=\linewidth]{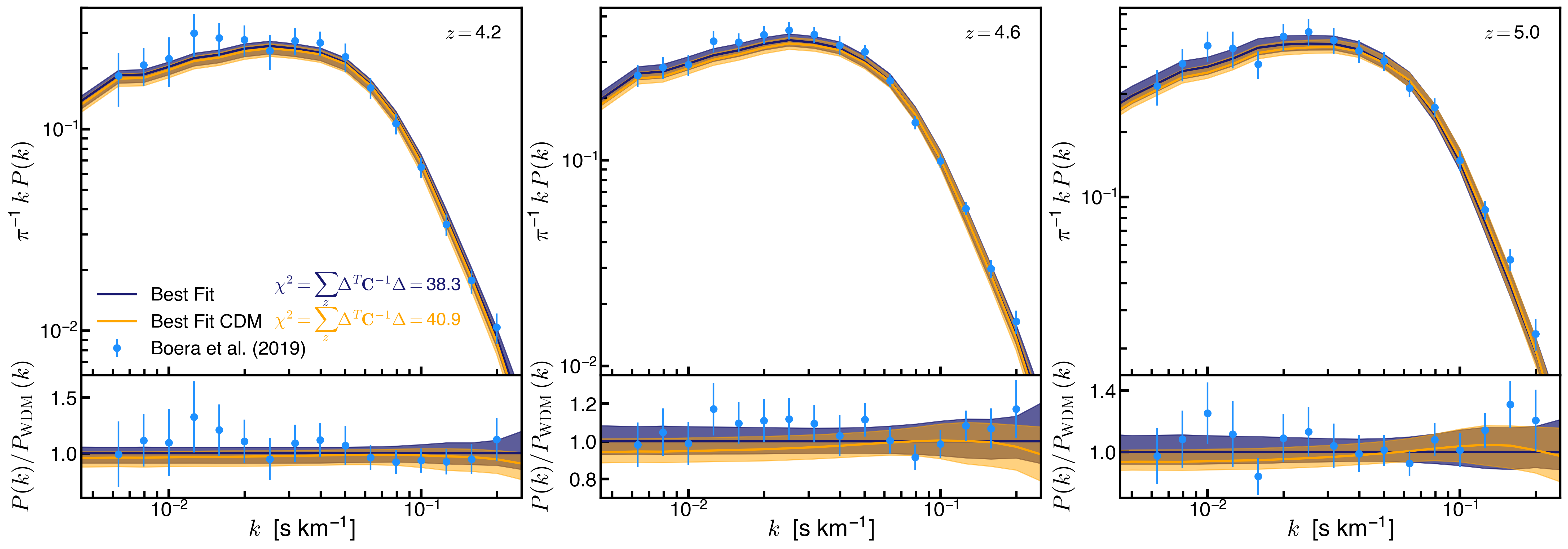}
    \caption{Observational determination of \pk\ in the redshift range $4.2 \leq z \leq 5.0$ used in this work to constrain WDM cosmologies. 
    The best-fit from the full WDM-CHIPS grid (purple) and the subset CDM grid (orange) are shown as lines and shaded regions which 
    correspond to 95\% confidence range of \pk\ marginalized over the posterior distribution.
    From our analysis, the \pk\ from WDM cosmologies 
    with $\mwdm = 4.5$ keV is preferred 
    over the best-fit CDM model. We note that the WDM best-fit results in lower IGM temperatures compared to CDM 
    (see Figure \ref{fig:temperature_best_fit}). The bottom panels show the ratios between the observations and the best-fit WDM FPS 
    (blue points) and the best-fit CDM and WDM  FPS (orange line), along with the WDM and CDM model 95\% confidence intervals (shaded regions). 
    }
    \label{fig:ps_best_fit}
\end{figure*}

The comparison presented in Figure \ref{fig:ps_real_space} suggests that peculiar velocities play an influential
role shaping the structure of the \Lya\ forest on small scales.
We conclude that for $k \gtrsim 0.1 \,\, \skm$ 
peculiar velocities are more important for shaping \pk\
than free-streaming.
Since peculiar velocities and thermal broadening on the smallest scales
affect WDM and CDM models
similarly, the decrease in \pk\ from WDM relative to CDM saturates at $k \gtrsim 0.1 \,\, \skm$.  

\section{Statistical Comparison}

In this section we present the observations of \pk\ used for this work and we describe the methodology used for the Bayesian approach used in this work to constrain 
the models used for our simulation suite.    

\begin{figure*}[htb!]
    \centering
    \includegraphics[width=\linewidth]{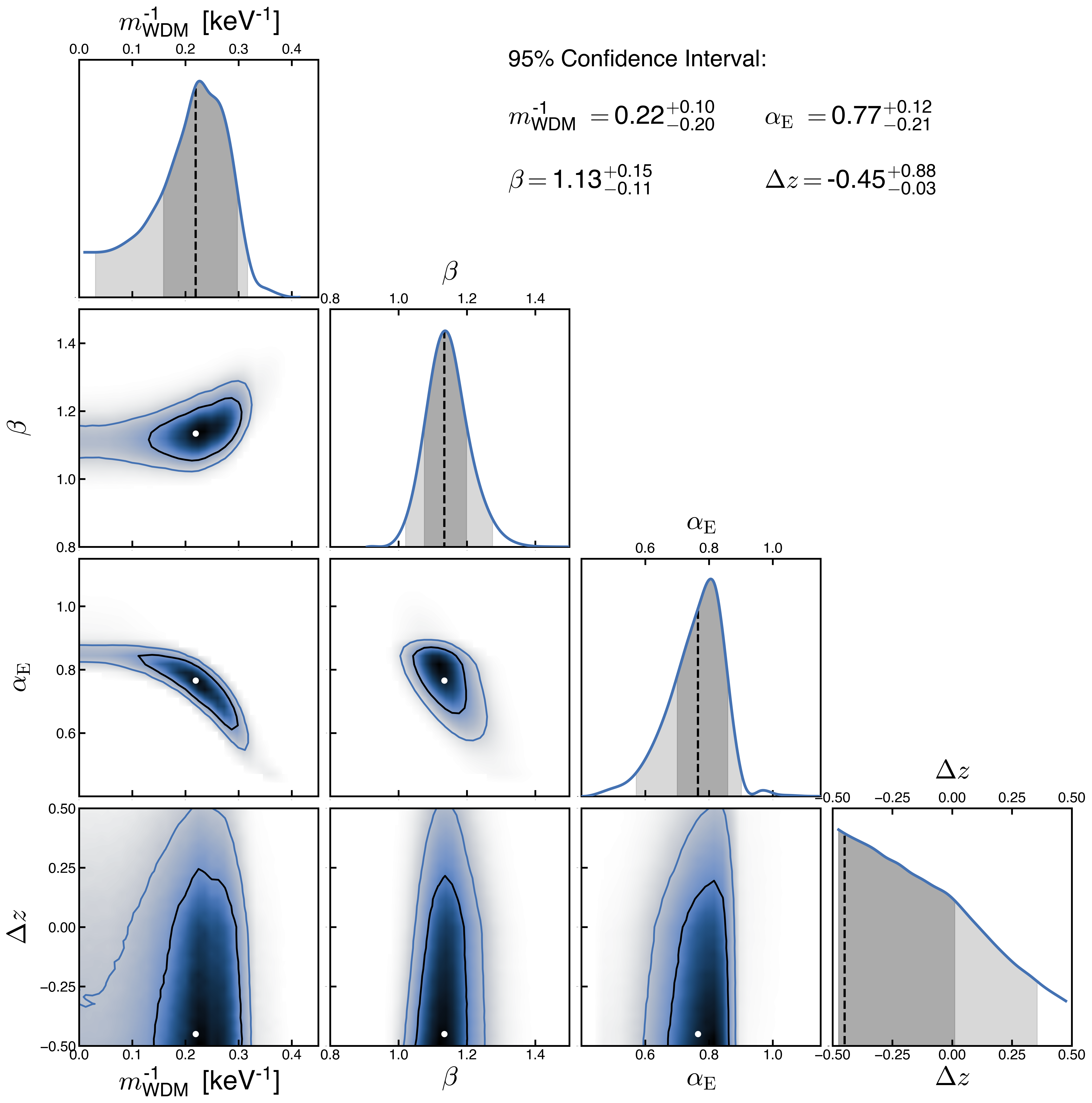}
    \caption{Results from the Bayesian inference procedure, showing one- and two-dimensional projections of the posterior
    distributions for the parameters $\theta= \{ m_{\rm WDM}^{-1}, \beta, \,\alpha_E,\,\Delta z \}$ 
    recovered by fitting synthetic flux power spectra from our grid of WDM-CHIPS simulations to observations of the \Lya\ forest from \cite{boera2019a}. Gray bands in the 1D distributions show the $1\sigma$ and $2\sigma$ intervals. The marginalized likelihood for $1/m_{\rm WDM}$ peaks at $1/(4.5\,{\rm keV})$. The preference for a non-vanishing free-streaming length is only weakly statistically significant, as the CDM case is contained within the $3\sigma$ interval of the distribution. Our main result is a lower bound $\mwdm > 3.1$ keV at the $2\sigma$ confidence level. 
    }
    \label{fig:corner_wdm}
\end{figure*}

\begin{figure*}[htb!]
    \centering
    \includegraphics[width=0.8\linewidth]{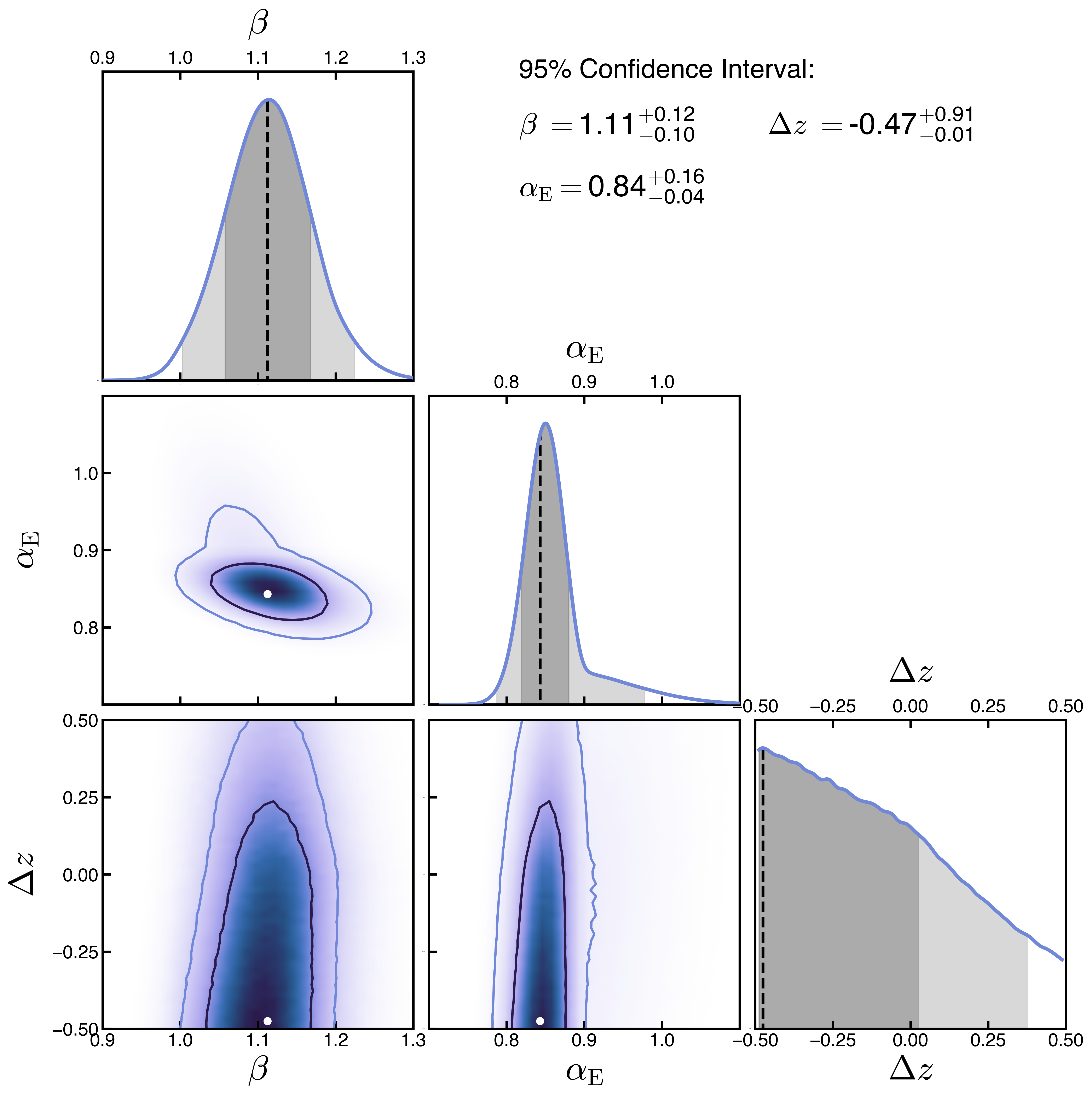}
    \caption{Results from our MCMC procedure, showing one- and two-dimensional projections of the posterior
    distributions for the UVB parameters $\theta= \{ \beta, \,\alpha_E,\,\Delta z \}$. Here the sampling of models is restricted to the subset of 90 simulations that evolve a CDM cosmology. Gray bands in the 1D distributions show the $1\sigma$ and $2\sigma$ intervals. 
    }
    \label{fig:corner_cdm}
\end{figure*}

\subsection{Observational Flux Power Spectrum }

For this work we employ observations of the \Lya\ forest power spectrum from the Keck observatory and the Very Large Telescope, presented in \cite{boera2019a}. 
These measurements represent the highest resolution determination of \pk\ to date and probe the structure of the forest in the redshift range $4.2 \leq z \leq 5.0$.
Since the impact of free-streaming on \pk\ is greater at high redshift
 the data set from \cite{boera2019a} is arguably the most constraining determination of \pk\ 
to infer the nature of dark matter currently available.
The measurements of \pk\ from \cite{boera2019a} along our best-fit determination obtained from our analysis are shown in Figure \ref{fig:ps_best_fit}.

\subsection{MCMC Inference}
\label{sec:mcmc_inference}

To constrain WDM cosmologies from the observations of \pk , we use an MCMC approach to sample over our suite of models and obtain best-fit distributions for our
parameters. The likelihood function for the model given by the parameters $\theta= \{m^{-1}_{\rm WDM}, \beta, \,\alpha_E,\,\Delta z\}$ is evaluated as 

\begin{equation}
\ln \mathcal{L}(\theta) =   - \frac{1}{2}  \sum_z \left[  \mathbf{\Delta}^{T} \mathbf{C}^{-1} \mathbf{\Delta} +  \ln \operatorname{det}(\mathbf{C}) + N \ln 2 \pi  \right],
\label{eq:mcmc_likelihood}
\end{equation}
\noindent
where $\mathbf{\Delta}$ denotes the difference vector between the observational \pk\ and the model $\mathbf{\Delta} = P_{\mathrm{obs}}(z,k) - P(z,k|\theta)$, and 
$\mathbf{C}$ corresponds to the covariance matrix associated with the observation of \pk\  as reported by \cite{boera2019a}.
These authors have considered how systematic uncertainties may affect their covariance matrix, and find that the level of potential systematics 
is $\lta 1\%$. To compute $P(z,k|\theta)$ for arbitrary 
values of the parameters $\theta$ not directly simulated by our grid, we perform a four-dimensional linear interpolation of the FPS measured from the sixteen neighboring simulations in parameter space.
The methodology for our 
inference approach differs from previous works in a number of key respects:

\begin{enumerate}[leftmargin=*]
\item Our WDM-CHIPS simulation grid captures a wide range of  free-streaming lengths and UVB models to sample over a variety of cosmic structure formation and gas thermal histories, and thereby produce different 
statistical properties for the \Lya\ forest. Due to the computational cost of the simulations, evolving 
such a large grid of high-resolution cosmological hydrodynamical simulations
had not been achieved before. With the advent 
of efficient numerical codes like {\sc Cholla}  and capable systems like Summit, it is now possible to simulate thousands of models.        
\item We evaluate the performance of a given model in matching the observations of \pk\ by marginalizing over the self-consistently evolved hydrogen reionization history 
in the redshift interval $4.2\leq z\leq 5.0$. The IGM thermal properties at one redshift cannot be disentangled from its properties at previous epochs, thus the marginalization over the parameter posterior distributions should not be performed independently at each redshift \cite[c.f.][]{bolton2014a, nasir16, hiss2018a,  boera2019a,   walther2019a, gaikwad2020b}).

\item We do not modify, in post-processing, the mean transmitted flux \Fmean\ in the forest by rescaling the \Lya\ optical depth, nor do we assume an instantaneous power-law temperature-density relation (cf. \cite{viel2013a, Irsic+2017b, boera2019a, walther2021}). 
The latter does not accurately reproduce the thermal state of IGM gas in the range $-1 \leq \log_{10} \Delta \leq 1$. Instead, in our approach simulations self-consistently evolve 
the ionization and thermal structure of the IGM determined by the wide range of photoionization and photoheating histories applied in our model grid.
\end{enumerate}

\begin{figure}[htb!]
    \centering
    \includegraphics[width=\linewidth]{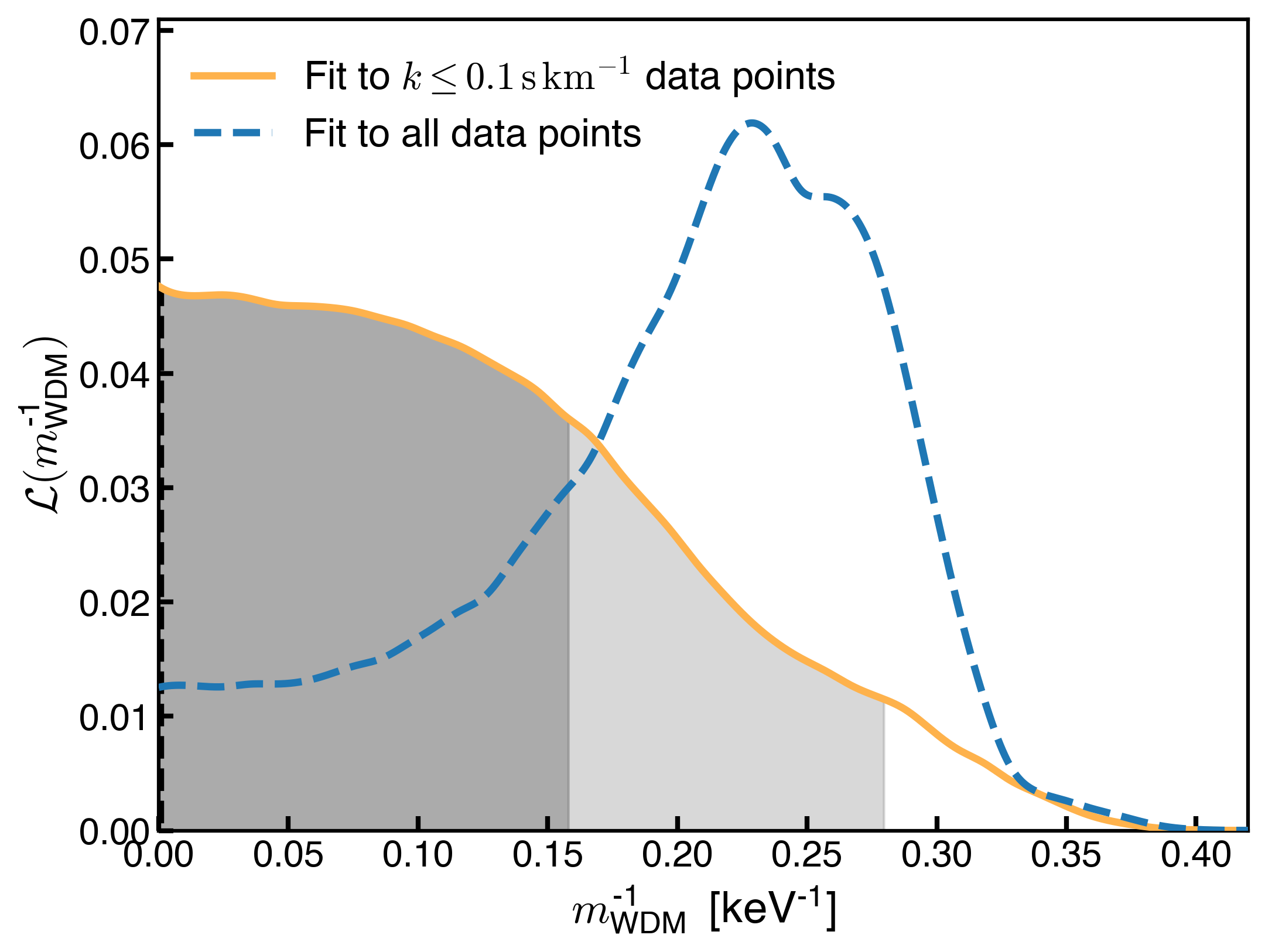} 
    \caption{One-dimensional posterior likelihood for the WDM particle mass obtained by limiting the observational measurements of \pk\ \cite{boera2019a} 
    to $k \lesssim 0.1 \, \skm$ compared to the likelihood distribution from fitting to 
    the complete data set (dashed blue). The limited data set for the observational \pk\ is constructed by excluding the last three points of each redshift snapshot. Gray bands show the $1\sigma$ and $2\sigma$ interval from the fit to the limited data. In this case $\Lambda$CDM is the 
    preferred cosmology and the lower limit at 95\% confidence shifts to $\mwdm > 3.6 \, \mathrm{keV}$ compared to $\mwdm > 3.0 \, \mathrm{keV}$ from
    the fit to the full data set. }
\label{fig:mwdm_limited_data}
\end{figure}

\begin{figure}[htb!]
    \centering
    \includegraphics[width=\linewidth]{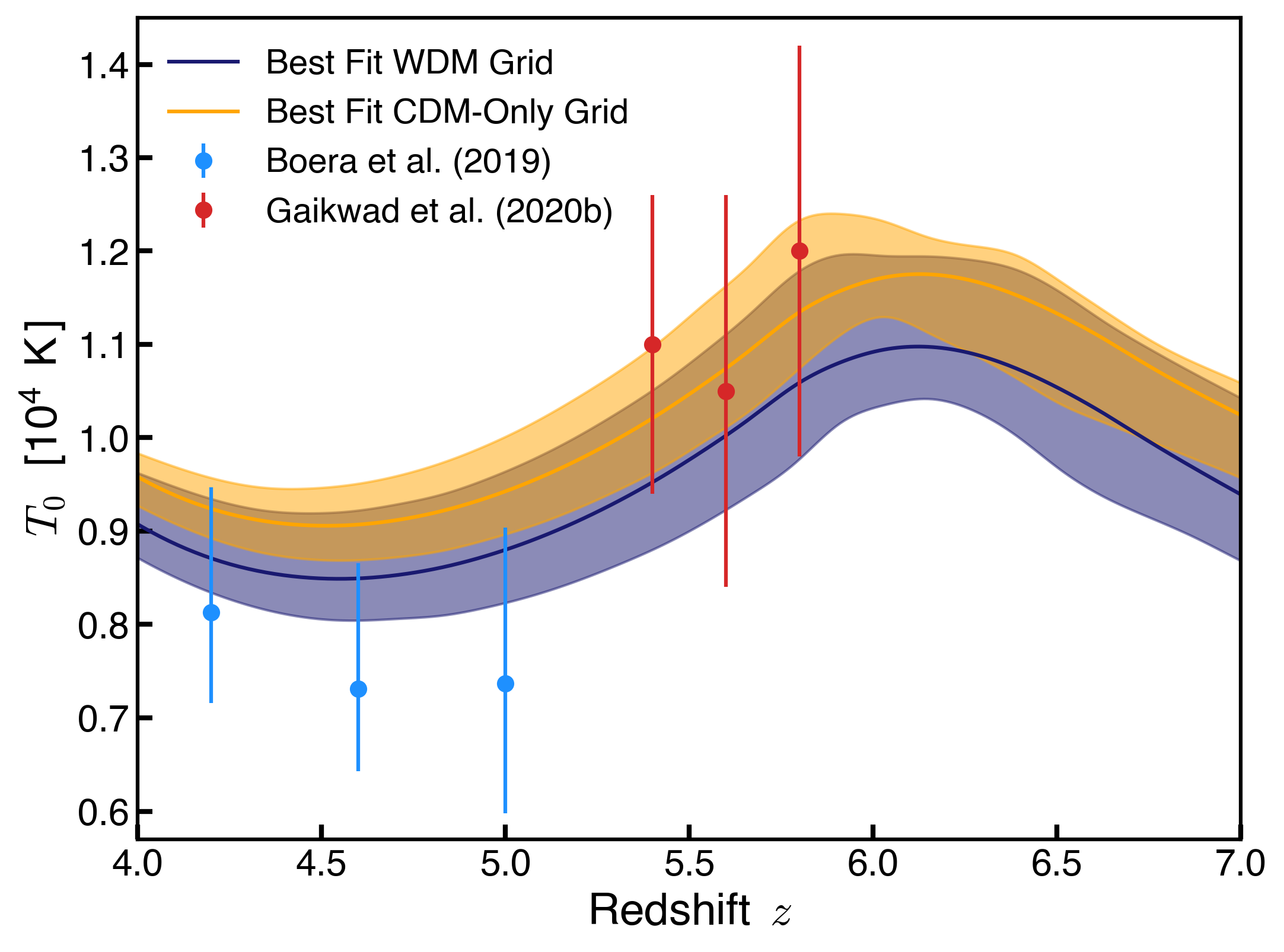} 
    \caption{Redshift evolution of the temperature $T_0$ of IGM gas at the mean density
    from the best-fit $\Lambda$CDM and WDM models (solid lines) and $1\sigma$ interval (colored bands) obtained from our MCMC analysis. 
    The data points show the values of  $T_0$ inferred from observations of the \Lya\ forest by \cite{boera2019a} and \cite{gaikwad2020a}.
    The best-fit from WDM cosmologies have moderately lower temperatures compared to CDM, which compensates for the suppression of small-scale
    \pk\ due to free-streaming.
    }
\label{fig:temperature_best_fit}
\end{figure}

\section{Results}

By comparing the high-resolution observation of the flux power spectrum from \cite{boera2019a} to our suite of simulations that simultaneously vary the impact of 
free-streaming from WDM cosmologies 
on the matter distribution and different reionization and thermal histories of the IGM, we infer new constraints on the WDM particle mass \mwdm . In this section we present the 
posterior distribution from our MCMC analysis and the marginalized \pk\ and thermal evolution of the IGM for our best-fit models. To complement our results, we repeat our analysis
by modifying \pk\ from the simulations to account for a non-uniform UV background, and finally, we show hypothetical constraints on \mwdm\ from an artificially increased number of quasars.

\subsection{Distribution of the Model Parameters}

The posterior distribution of our four model parameters $\theta= \{ m_{\rm WDM}^{-1}, \beta, \,\alpha_E,\,\Delta z \}$ resulting from the Bayesian inference procedure is 
shown in Figure \ref{fig:corner_wdm}. A clear global maximum of the 1D marginalized distributions is shown for \mwdm\ and the parameters $\beta$ and \alphae\ responsible
for rescaling the photoionization and photoheating rates. The distribution of the parameter $\Delta z$ is not fully contained by our grid of models, but we 
argue that this issue does not represent a significant challenge to our conclusions. Values of $\Delta z < -0.5$ would reduce the suppression of the small-scale \pk\ 
from pressure smoothing which could be compensated by free-streaming or thermal broadening, possibly shifting our constraint of \mwdm\ to slightly lower values.

The marginalized distribution for $m_\mathrm{WDM}^{-1}$ (top left panel, Figure \ref{fig:corner_wdm}) is well constrained and it peaks at $m_\mathrm{WDM}^{-1} = 1 /(4.5 \, \, \mathrm{keV})$. Arguably, the preference for a non-zero free-streaming length is weakly significant as the CDM cosmology is contained within the $3\sigma$ interval of the marginalized distribution. The principal result from our analysis is the lower limit $\mwdm > 3.0 \, \mathrm{keV}$ (at the $2\sigma$ level) obtained for 
the WDM particle mass. We have also performed our MCMC analysis sampling in $m_{\rm WDM}$ space (instead of $m_{\rm WDM}^{-1}$ space) applying a flat prior in mass the mass range
of $1<m_{\rm WDM}<80$ keV, and we find that our conclusions are
unchanged. We also note that the sensitivity of the best-fit WDM mass on the UVB parameters (and hence on the reionization history) is
largely a thermal Doppler broadening (and to a less extent, a pressure) effect.

To find the best-fit model to \pk\ for a CDM cosmology, we repeat our MCMC approach but restricted to the models that evolve the 
$\Lambda$CDM cosmology varying only the photoionization and photoheating history. Figure \ref{fig:corner_cdm} shows the one- and two-dimensional 
projections of the posterior distribution for the UV background (UVB) parameters $\{ \beta, \,\alpha_E,\,\Delta z \}$ obtained from the comparison of \pk\ from \cite{boera2019a}
to the grid of 90 CDM simulations. 
The posterior distribution of the UVB parameters from our CDM fit differs slightly from the distribution of parameters 
obtained from the inference that included WDM cosmologies (see Figure \ref{fig:corner_wdm}). The main difference between the two distributions
is that the parameters $\beta$ and \alphae\ peak at slightly shifted values. In particular, the product $\beta \, \alphae$, which rescales the photoheating rates, is higher 
for the CDM case. Higher photoheating rates directly result in increased IGM temperatures (see Section \ref{sec:temperature_evolution}).

\subsection{Best-Fit Power Spectrum}

The marginalized flux power spectrum \pk\ over the posterior distribution obtained from our MCMC analysis along 
the observational data used to constrain the model is shown in Figure \ref{fig:ps_best_fit}. Lines and shaded regions 
show the best-fit \pk\ and the 95\% confidence range. The result from the fit performed by sampling over the entire WDM model grid (purple)
and restricting to the CDM grid (orange) are shown separately. As shown, both results provide a good match to the observed \pk\, nevertheless, 
the relative difference between the model and the data quantified as $\chi^2= \sum_z  \Delta^{T} \mathbf{C}^{-1} \Delta$ is slightly lower for the best-fit 
WDM model ($\chi^2= 38.3$), compared to $\chi^2= 40.9$ from the CDM best-fit model. Here $\mathbf{C}$ denotes the covariance matrix taken from \cite{boera2019a}
and $\Delta$ is the difference vector between the observed and model \pk . 

We find that the preference for a non-zero free-streaming scale derived from our analysis is driven by the smallest scales probed by the high-resolution 
measurement of \pk\ from \cite{boera2019a}. 
We repeated our inference methodology but excluding the three last data points ($k > 0.1 \,\, \skm$) from the likelihood calculation. Here we
found that excluding the high-$k$ measurements places $\Lambda$CDM as the preferred cosmology. 
The marginalized posterior distributions for \mwdm\ from the fits limited to 
$k \leq 0.1 \,\, \skm$ and to full data set are displayed in Figure \ref{fig:mwdm_limited_data}. As shown, the likelihood peaks at CDM and the lower limit constraint 
at the $2\sigma$ level shifts to $\mwdm > 3.6 \, \mathrm{keV}$.

In section \ref{sec:pk_variation} we showed that free-streaming and Doppler broadening impact \pk\ differently on small scales ($k > 0.1 \, \skm$). As the temperature of 
the gas is increased/decreased, the impact of thermal broadening is to decrease/increase the small-scale \pk\ .
While free-streaming decreases small-scale density fluctuations, its effect on the flux power spectrum saturates at $k \gtrsim 0.1 \, \skm$ 
due to the impact from peculiar velocities 
(see Figs. \ref{fig:ps_effects_comparison}, \ref{fig:ps_real_space}).

We have shown that the preference for a WDM cosmology over $\Lambda$CDM 
originates from the $k> 0.1 \, \skm$ measurement of \pk . Also, we showed that on these scales ($k> 0.1 \, \skm$) the suppression of \pk\ due to free-streaming saturates
while the suppression of \pk\ from thermal broadening increases monotonically. Therefore, we conclude that the saturated suppression 
of \pk\ due to a $\mwdm \sim 4.5 \, \mathrm{keV}$ combined with a moderately lower IGM temperature 
provides a slightly better fit to the $k> 0.1 \, \skm$ observation of \pk\ compared to the $\Lambda$CDM best-fit model with
higher IGM temperatures at $4 \leq z \leq 5$ (see Fig. \ref{fig:temperature_best_fit}).

\subsection{Thermal Evolution of the IGM}
\label{sec:temperature_evolution}

One of the advantages of our approach compared to previous works that aim to constrain the WDM free-streaming from observations of the \Lya\ forest power spectrum is that
our simulations self-consistently evolve the IGM thermal and ionization history during and after reionization by sampling over a large grid models that vary the UVB photoionization and 
photoheating rates. Instead, the methodology adopted by previous works \cite{viel2013a, Irsic+2017b, garzilli2019, garzilli2021} was to change the temperature $T_0$ and the ionization fraction of the IGM in post-processing by rescaling $T_0$ and the effective optical depth \taueff of the simulated skewers.

Figure \ref{fig:temperature_best_fit} shows the redshift evolution of the temperature of the gas at mean density $T_0$ marginalized over the posterior distribution
from our MCMC analysis; shaded bars show the $1\sigma$ interval. The best-fit model for WDM cosmologies with $\mwdm=4.5\,$keV results in moderately lower ($5-10$\%) 
IGM temperatures (purple)
due to the slightly reduced photoheating rates compared to the $\Lambda$CDM best-fit model (orange). Lower temperatures are expected from the WDM models as they decrease the impact of
thermal broadening in suppressing \pk\ which compensates 
for the suppression due to free-streaming.

In Figure \ref{fig:temperature_best_fit}, we compare the IGM temperature $T_0$ inferred by this work to other measurements at $z>4$ \cite{boera2019a, gaikwad2020a}. 
The high-redshift inference presented by \cite{gaikwad2020a} was obtained by characterizing the transmission spikes observed in $z>5$ spectra. 
The thermal histories obtained from our analysis are consistent with the results from \cite{gaikwad2020a}, suggesting a peak in $T_0$ due to hydrogen reonization at $z \sim 6$.

The measurement of $T_0$ presented in \cite{boera2019a} for $4.2 \leq z \leq 5.0$ was made by fitting the observed \pk\ to simulated spectra according to a $\Lambda$CDM cosmology
where the instantaneous density-temperature distribution of the gas is modified in post-processing changing parameters $T_0$ and $\gamma$ from the power-law relation 
$T= T_0 (\rho/\bar{\rho})^{\gamma-1}$. Despite using the same observational determination of \pk\ for our inference, the IGM temperatures obtained in this work during 
$4.2 \leq z \leq 5.0$ are slightly higher than those inferred by \cite{boera2019a}.
Restricting to a $\Lambda$CDM cosmology, we find a best-fit $T_0$ is $15 - 20$\% higher
at redshift $z=5.0$ and $z=4.6$ compared to the result from \cite{boera2019a}. At redshift $z=4.2$ the difference lowers to 10 - 15\% and our result agrees with their inference within $1\sigma$. The evolution of $T_0$ from our best-fit WDM model is also higher (5 - 10\%) than the results from \cite{boera2019a}, but in this 
case our results also agree to within $1\sigma$. 

\subsection{Thomson Scattering Optical Depth}

\begin{figure}[htb!]
    \centering
    \includegraphics[width=\linewidth]{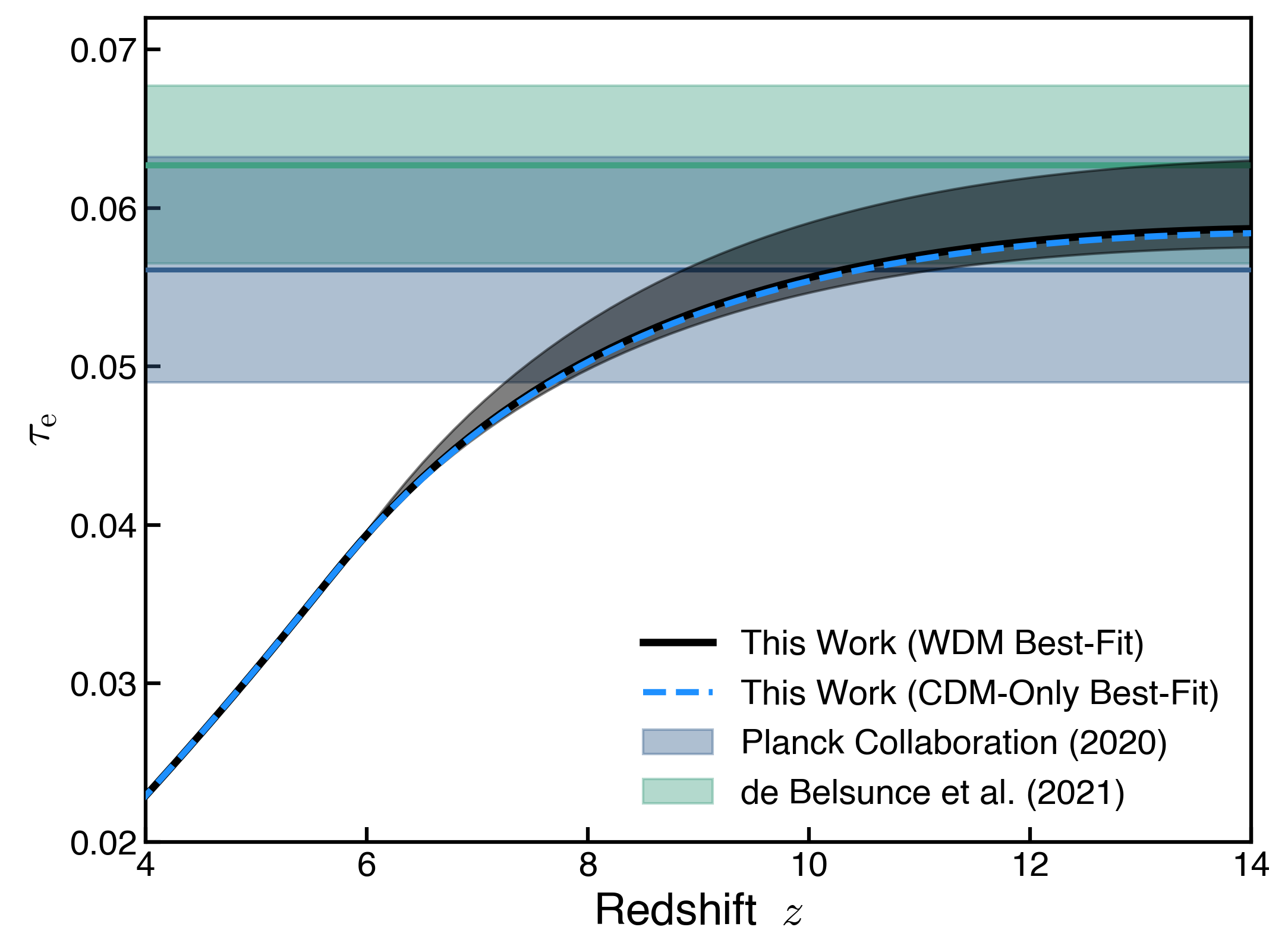} 
    \caption{Electron scattering optical depth, $\tau_e$, 
    to reionization. Black line and shaded bar: WDM best-fit model and 1$\sigma$ interval. Dashed blue line: CDM-Only best-fit model. Also shown are the observational constraints from the Planck satellite
\cite{Planck_collaboration_2020,deBelsunce+2021}.
    } 
\label{fig:tau_electron}
\end{figure}

Our WDM and CDM-Only best-fit models result in very similar reionization histories with Thomson scattering optical depths that are consistent with constraints from the Planck  satellite \cite{Planck_collaboration_2020,deBelsunce+2021}. Figure \ref{fig:tau_electron} shows such constraints together with the marginalized Thomson scattering optical depth, $\tau_\mathrm{e}$, for our WDM best-fit model (black line,
with the shaded regions displaying the 1$\sigma$ interval), and for our CDM-Only best-fit model (blue line). We measure $\tau_\mathrm{e}=0.0586^{+0.0043}_{-0.0012}$ and $\tau_\mathrm{e}=0.0584^{+0.0041}_{-0.0010}$ from our WDM and CDM-Only best-fit simulations, respectively.

\begin{figure}[htb!]
    \centering
    \includegraphics[width=\linewidth]{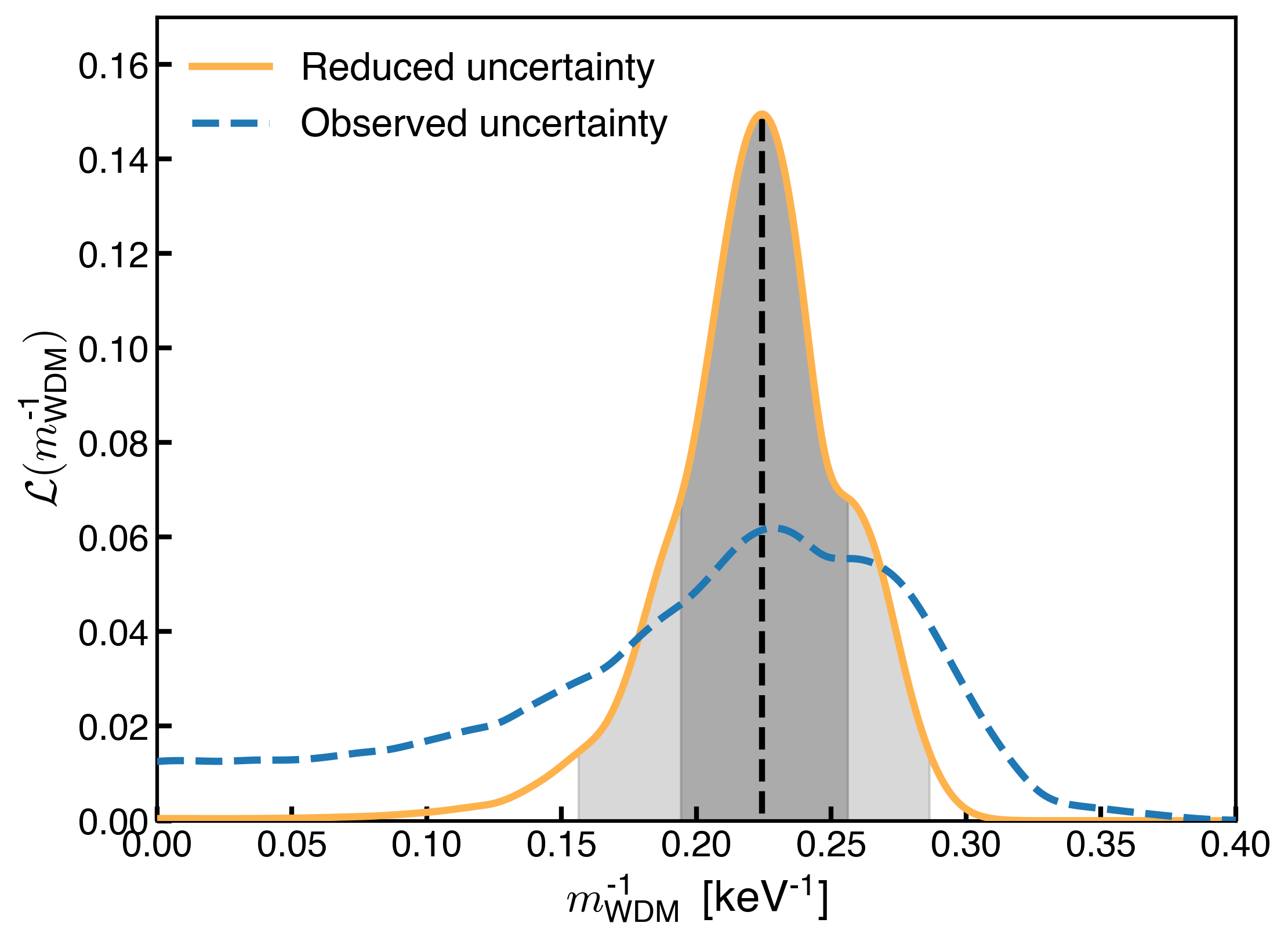} 
    \caption{One-dimensional posterior likelihood for the WDM particle mass obtained by fitting to the measurements of the FPS from \cite{boera2019a} 
    but with an artificially reduced uncertainty motivated by the improved statistics of \Lya\ spectra from upcoming surveys. Here we rescale 
    the covariance matrix $\mathbf{C}$ of \pk\ by a factor of one-fourth. In this hypothetical case, the constraint on \mwdm\ is tighter, 
    measured as $\mwdm = 4.5_{-1.0}^{+1.9} \, \mathrm{keV}$ at 95\% confidence level. The dashed line shows our result
    from fitting to \pk\ with the reported uncertainty from \cite{boera2019a}.
    } 
\label{fig:mwdm_reduced_sigma}
\end{figure}

\begin{figure*}[htb!]
    \centering
    \includegraphics[width=\linewidth]{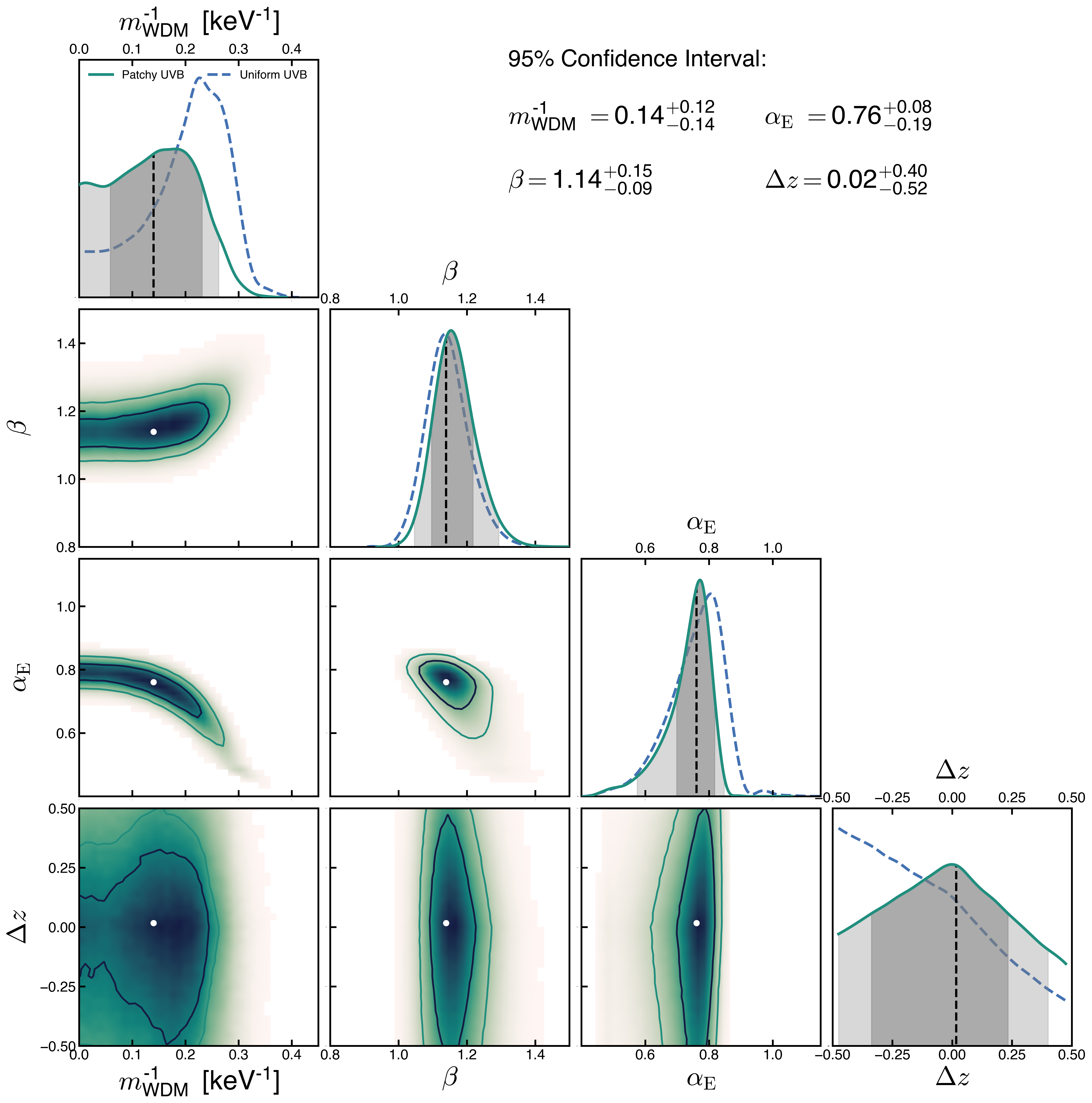}
    \caption{
    Posterior distribution of the parameters $\theta= \{ m_{\rm WDM}^{-1}, \beta, \,\alpha_E,\,\Delta z \}$ from fitting the observed \pk\ from \cite{boera2019a} 
    with models that account for a patchy reionization according to the modification to \pk\ presented in \cite{Molaro+2022}. In this case, 
    the preference for a WDM cosmology persists but the distribution shifts to higher mass. The likelihood peaks at $\mwdm = 7.1 \, \mathrm{keV}$ and the lower limit at the 
    95\% levels is $\mwdm = 3.8 \, \mathrm{keV}$. In this case the $\Lambda$CDM cosmology is contained within the 95\% interval of the distribution. The dashed blue
    lines show the one-dimensional likelihood distributions obtained from fitting \pk\ measured directly from the uniform UVB simulations. 
    }
    \label{fig:corner_wdm_inhomogeneous}
\end{figure*}

\subsection{Constraining WDM with Increased Quasar Sightlines}

Upcoming surveys of the \Lya\ forest (e.g., DESI \cite{desi2016a}, WEAVE \cite{Pieri+2016}, EUCLID \cite{Amiaux+2012}, LSST) will drastically increase the available 
observations of quasar sightlines, which will significantly improve the statistics of measurements derived from the forest. The improved statistics will allow tighter constraints on cosmological parameters as well as on WDM and the sum of the neutrino masses from their role in suppressing small-scale structure in the forest.

To assess how better statistics would impact the constraining power of high-redshift and high-resolution observations of the \Lya\ forest for WDM cosmologies, we repeat our analysis but
decreasing the uncertainty on the observed \pk\ by a factor of one half, which would correspond to increasing the number of
observed quasar spectra from fifteen used by \cite{boera2019a} for their measurement of \pk , to about sixty (a factor of four increase).

In our approach, we approximate a more constraining dataset by rescaling the covariance matrix $\mathbf{C}$ of \pk\ by a factor of one-fourth in Equation  (\ref{eq:mcmc_likelihood}).  
Note that the observational measurements of the FPS  are not altered, and only the covariance matrix is reduced. 
Figure \ref{fig:mwdm_reduced_sigma} shows the marginalized likelihood for $\mwdm^{-1}$ obtained from our analysis using the reduced uncertainty. In this hypothetical case, the 
improved statistics of the \pk\ measurement provide a tighter constraint for the WDM particle mass measured as $\mwdm = 4.5_{-1.0}^{+1.9} \, \mathrm{keV}$ at the 95\% confidence level.
This exercise demonstrates that increasing the sample of high-$z$ and high-resolution observations of the \Lya\ forest should place tight constraints on WDM cosmologies. 
 
\subsection{Modified $P(k)$ for Inhomogeneous UVB }

One of the main limitations of our analysis
is that the WDM-CHIPS simulations evolve under a homogeneous UVB in the form of uniform photoionization and photoheating rates.   
We evaluate the effect of this assumption on our conclusions by repeating our analysis but using the modification to \pk\ prescribed
by \cite{Molaro+2022} to account for the impact of a nonuniform UVB on the simulated \pk .

In \cite{Molaro+2022}, a set of simulations that apply a uniform UVB were compared to a set that follows a hybrid-RT method. 
For the latter, spatially-varying maps for the \HI\ photoionization and photoheating rates were computed in post-processing 
and used as input for a re-run of the base simulation, incorporating the response of the gas to the non-uniform photoionization and photoheating. For the comparison, pairs 
of uniform and nonuniform UVB simulations were calibrated to have the same average ionization and thermal history. The authors concluded that simulations with a patchy reionization show a suppression ($10-15$\%) of the FPS on small scales ($k\sim 0.1\,\skm$) with 
respect to the uniform UVB case. This effect is mainly driven by the Doppler broadening associated with the high temperatures of recently ionized regions and the divergent 
peculiar velocities of thermally pressurized gas \cite{Xiaohan+2019}. On large scales ($k > 0.03 \, \skm$) the variation of the IGM neutral fraction due to large-scale fluctuations of the gas temperature leads to an increase on \pk. 

The likelihood distribution from our MCMC inference where we modify the \pk\ from each one of the simulations in our WDM-CHIPS grid to account for an 
inhomogeneous reionization according to the transformation presented in \cite{Molaro+2022} is shown in Figure \ref{fig:corner_wdm_inhomogeneous}. Dashed blue lines correspond to 
the one-dimensional marginalized distributions obtained from fitting the base \pk\ from the simulations with a uniform reionization. 

Figure \ref{fig:corner_wdm_inhomogeneous} shows that modifying the models \pk\ to account for patchy reionization mainly affects the likelihood of \mwdm. 
While WDM cosmologies are still preferred over $\Lambda$CDM when fitting the non-uniform UVB models, the likelihood shifts to higher values of \mwdm\ compared to the result 
from uniform UVB models. An inclination for models with reduced free-streaming compensates for the reduction of small-scale \pk\ due to patchy reionization. In this case the 
maximum likelihood occurs at $\mwdm = 7.1 \, \mathrm{keV}$ and the lower limit at 95\% level is at $\mwdm =  3.8 \, \mathrm{keV}$. Notably, for the patchy 
UVB models, the $\Lambda$CDM cosmology is contained within the 95\% interval of the likelihood distribution. 

We note that the difference between the observed and model \pk\ defined as $\chi ^2 = \sum_z   \mathbf{\Delta}^{T} \mathbf{C}^{-1} \mathbf{\Delta} $ 
(see \S\,\ref{sec:mcmc_inference} is larger for the nonuniform UVB modified best-fit models compared to the best-fit models sampling from the uniform UVB simulated \pk .
When fitting with \pk\ measured directly from the uniform UVB simulations, we measure $\chi^2 = 38.3 $ for the best-fit model with $\mwdm = 4.5 \, \mathrm{keV}$ 
and a slightly higher $\chi^2 = 40.9 $ for the $\Lambda$CDM best-fit model.
On the other hand, when sampling over the modified \pk\ to account for patchy reionization, we measure larger values $\chi^2 = 46.4$ and $\chi^2 = 46.6$ for the 
WDM and CDM best-fit models, respectively.
The higher values of $\chi^2$ obtained for the patchy reionization modified models show that, in the context of this work, \pk\ measured directly from the uniform UVB simulations provide a better match to the 
observation from \cite{boera2019a}, therefore we use the likelihood distribution obtained from the sampling over the uniform UVB models to construct the main results from our analysis.  
     
\section{Conclusions}

To constrain 
cosmological models where dark matter free-streaming smooths the matter distribution in the Universe, we have used the GPU-native code {\sc Cholla} 
to perform a massive suite of high-resolution hydrodynamical cosmological simulations that simultaneously vary the effect of free-streaming from WDM particles and 
the IGM reionization history.
We compare the power spectrum of the synthetic \Lya\ forest from our simulations to the high-resolution observational measurement presented by \cite{boera2019a} to determine via a 
likelihood analysis the optimal model for cosmological free-streaming that best matches the observation. A summary of the efforts and conclusions from this work follows.

\begin{itemize}

\item We present the WDM-CHIPS suite consisting of a grid of 1080 high-resolution simulations ($ L = 25 h^{-1} \mathrm{Mpc}$, $N=2\times1024^3$) that vary the free-streaming from WDM
cosmologies and the photoionization and photoheating rates from the metagalactic UVB. The UVB rates applied for our grid use the model from \cite{villasenor2022} as a template,
and use three parameters that control a rescaling amplitude and redshift-timing of the hydrogen photoionization and photoheating rates. Combined with the WDM particle mass \mwdm ,
our four-dimensional grid of models densely sample a wide range of self-consistently evolved reionization and thermal histories of the IGM. This represents a significant improvement 
over previous studies that aimed to constrain WDM cosmologies from observations of the \Lya\ forest  \cite{viel2005, viel2013a, Irsic+2017b, garzilli2019, garzilli2021}. 

\item The large range of thermal histories produced by the different UVB models in our grid of simulations results in synthetic measurements of \Lya\ spectra where the 
impact from Doppler broadening and pressure smoothing on suppressing the small-scale \pk\ varies widely.
This flexibility is important as these mechanisms have similar effects
as free-streaming decreasing small-scale fluctuations in the \Lya\ forest. Additionally, our approach does not require an assumption of a power-law relation 
for the density-temperature distribution of the gas.
Self-consistent evolution of the IGM phase structure
proves to be important as we find that a single power law does not accurately describe the $\rho_\mathrm{gas}-T$ 
distribution in the density range relevant to generating the signal of the \Lya\ forest (see Appendix E of \cite{villasenor2022}).   

\item We compare our grid of models to the high-redshift ($4.2 \leq z \leq 5.0$) observational measurement of the \Lya\ forest power spectrum from \cite{boera2019a}. 
The observations presented in \cite{boera2019a} provide the highest resolution measurement 
of \pk\ at the time that work was executed. We perform a Bayesian MCMC sampling to determine the best-fit model for free-streaming 
due to WDM cosmologies and the IGM photoionization and photoheating history. 

\item From our MCMC analysis, we find that a cosmological model with non-vanishing free-streaming is preferred over $\Lambda$CDM. 
The likelihood function peaks at the thermal relic mass of $\mwdm = 4.5 \,\, \mathrm{keV}$ and a we measure a lower limit of $\mwdm = 3.1 \,\, \mathrm{keV}$ at the 95\% CL.
We find a weak ($3\sigma$) preference for WDM over $\Lambda$CDM, but both are statistically 
consistent with the currently available data.

\item
We repeat our MCMC analysis, restricting to $\Lambda$CDM only but with a variable UVB. We find that the best-fit $\Lambda$CDM model 
mainly differs from the WDM optimal model in that the IGM temperatures are slightly lower ($5-10$\%) for the WDM case. We find that both thermal histories are
in good agreement with the inference from \cite{gaikwad2020a} at $5.4 \lesssim z \lesssim 5.8$ and moderately higher than the temperatures from \cite{boera2019a} at 
$4.2 \leq z \leq 5.0$. 

\item We introduce the effect of a patchy reionization in our models by modifying \pk\ from our uniform UVB simulations according to the 
prescription presented in \cite{Molaro+2022}.
The likelihood distribution from our MCMC approach using the modified \pk\ shows that the preference for a WDM cosmology is maintained 
but the distribution shifts to higher values 
of \mwdm . The maximum likelihood and the 95\% lower limit constraint are $\mwdm= 7.1 \, \mathrm{keV}$ and $\mwdm= 3.8 \, \mathrm{keV}$, respectively. 
Additionally,
the $\Lambda$CDM case is $2\sigma$ consistent with the best-fit WDM model when accounting for an inhomogeneous UVB. Notably, the different $\chi^2$ between the observed and best-fit \pk\ model 
is higher for the modified model to account for a nonuniform UVB, and we therefore use the likelihood distribution obtained from the uniform UVB models as the basis for our results. 
 
\end{itemize}
\bigskip
\acknowledgments

This research used resources of the Oak Ridge Leadership Computing Facility at the Oak Ridge National Laboratory, which is supported by 
the Office of Science of the U.S. Department of Energy under Contract DE-AC05-00OR22725, using Summit allocations AST169 and AST175.  
An award of computer time was provided by the INCITE program, via project AST175.  We acknowledge use of the {\it lux} supercomputer at UC Santa Cruz, funded by NSF MRI grant AST1828315, and support from 
NASA TCAN grant 80NSSC21K0271. B.V. is supported in part by the UC MEXUS-CONACyT doctoral fellowship. B.E.R. acknowledges 
support from NASA contract NNG16PJ25C and grants 80NSSC18K0563 and 80NSSC22K0814. E.E.S. acknowledges support from NASA grant 80NSSC22K0720 and STScI grant HST-AR-16633.001-A. We wish to thank 
Nick Gnedin and Avery Meiksin for many 
useful inputs and valuable discussions.


%

\end{document}